\documentclass[a4paper,11pt]{article}
\usepackage{amsmath}
\def\euv{\varepsilon_{UV}}
\def\eir{\varepsilon_{IR}}
\def\hyg{{~_2F_1}}
\def\s{{\tilde{s}}}
\def\t{{\tilde{t}}}
\def\tb{{\bar{t}}}
\def\u{{\tilde{u}}}
\def\tpo{{\tilde{p_1}^2}}

\def\tps{{\tilde{p_3}^2}}

\def\xiwt{{\tilde{\xi}_w}}
\def\Li{{\rm Li}}
\def\Io{{\cal I}^{\rm (1)}}
\def\It{{\cal I}^{\rm (2)}_l}
\def\Ct{{\cal C}^{\rm (-2)}}
\def\Co{{\cal C}^{\rm (-1)}}
\def\Cz{{\cal C}^{\rm (0)}}
\begin{document}
\begin{flushright}
KEK-CP-160 \\
April,~2005
\end{flushright}
\vskip 2cm
\begin{center}
{\Large 
Dimensionally regularized one-loop tensor-integrals \\
with massless internal particles
}
\end{center}
\vskip 1.5cm
\begin{center}
{\Large Y. Kurihara} \\
\vskip 0.5cm
{\it High Energy Accelerator Research Organization,\\
Oho 1-1, Tsukuba, Ibaraki 305-0801, Japan}\\
\end{center}
\vskip 3cm
\begin{abstract}
A set of one-loop vertex and box tensor-integrals with massless internal
particles
has been obtained directly without any reduction method to scalar-integrals.
The results with one or two 
massive external lines for the vertex 
integral 
and zero or one massive external lines for the box integral
are shown in this report. 
Dimensional regularization 
is employed to treat any soft and collinear (IR) divergence.
A series expansion of tensor-integrals with respect to
an extra space-time dimension for the dimensional regularization
is also given.
The results are expressed by very short formulas in  
a suitable manner for a numerical calculation.
\end{abstract}
\newpage
\section{Introduction}
\renewcommand{\theequation}{\thesection.\arabic{equation}}
\setcounter{equation}{0}
The LHC (Large Hadron Collider) project\cite{LHC} at CERN is planned as the
next-generation energy-frontier experiment. 
One of the physics motivations of LHC experiments is to discover the Higgs
particle, which is the only one missing ingredient 
in the standard model.
In order to establish the model completely, it is essential to
find it and to investigate in its details. It should also be important
to search for new phenomena beyond the standard model
through any tiny deviation in experimental observations
from theoretical predictions. Further, not only searching for new
phenomena, but also performing
precise measurements of parameters included in the standard model is
another important issue of LHC. 

LHC has employed colliding proton-proton beams in order to achieve
beam energies as high as possible, which should enhance the possibility
to find new particles/phenomena.
However, a proton machine must have a 
large QCD background, since the proton is a
composite particle constructed by strongly interacting particles,
such as quarks and gluons. This veils 
signals with large backgrounds. In order to extract
as much physics information as possible from experimental data contaminated by
huge backgrounds, the behavior of the background should be understood
in detail. This implies that one should precisely understand QCD, because it
entirely governs the background.

The background of any proton-proton colliding experiment
must be completely predicted by QCD, in principle.
However, in fact, it is very a difficult task to make precise predictions 
because of the large coupling constant of QCD. 
Moreover the lowest level
(tree-level) calculation does not have any predictive power for 
the event rate, because there is no good
renormalization point well-defined experimentally. 
We need higher order perturbation 
calculations for precise predictions of the background behavior.

Loop integration is one of the critical issues 
for computing
these higher order corrections.
In general, N-point tensor- and scalar-integrals with massless
internal lines including infrared (IR) divergence must be calculated
in QCD.
Since an arbitrary number of dimensions
must be used in QCD to regulate any IR divergence,
the usual method for the standard model\cite{loop1} cannot be used directly.
Various methods to reduce ($N\geq5$)-point integrals into ($N-1$)-point
integrals\cite{reduction} with a dimensionally regulated 
scheme are proposed. Then, all of the necessary ($N\geq5$)-point integrals
can be expressed by a linear combination of box (four-point) 
tensor-integrals. Usually, box tensor-integrals are further reduced to 
four or less point scalar-integrals, and then numerically evaluated
to obtain higher order corrections. The IR finite box integrals are
obtained in \cite{loop2}; for the IR divergent case, box integrals 
with zero or one external mass are given in \cite{loop4}. 
All IR divergent box integrals are treated in \cite{loop5} using the partial
differential equation method.
Another approach to all possible box scalar-integrals
with massless internal lines with zero- to four-external massive lines 
is proposed
for the IR divergent case\cite{d&n} and the IR finite case\cite{d&n2}.
Recently, two independent formalisms were proposed 
for calculating one-loop virtual corrections
with an arbitrary number of external legs\cite{passarino,loop6}.

We propose a new method to calculate tensor-integrals directly, and not to use
a reduction method from tensor-integrals to scalar ones in this report.
A set of one-loop vertex and box tensor-integrals with massless internal
particles is given in terms of hypergeometric functions.
Dimensional regularization 
is employed to treat any IR-divergence. 
A series expansion of tensor-integrals with respect to
an extra space-time dimension in the dimensional regularization
is also given in this report.
All results are expressed by very short formulas with  
a suitable manner for numerical calculations.

A general form of three-point tensor-integrals is given in section 2, 
and that of four-point ones in section 3. The obtained results are
numerically checked in section 4. A numerical calculation method for the
hypergeometric function and their series expansion with respect to
an extra space-time dimension is given in Appendix $A$. 
The series expansion of the general form of 
tensor-integrals can be represented in terms
given in Appendix $A$. Those results are given in Appendix $B$.

\section{Vertex integral}
\setcounter{equation}{0}
\subsection{Massless one-loop {\it vertex} integral in n-dimension}
The tensor-integral of a massless one-loop vertex with rank $M \leq 3$ in a space-time
dimension of $n=4-2\euv$ can be written as
\begin{eqnarray*}
T_{\underbrace{\mu \cdots \nu}_{M}}^{(3)}&=&\left(\mu_R^2\right)^{\euv} 
\int  \frac{d^nk}{(2 \pi)^n}
\frac{k_{\mu} \cdots k_{\nu}}{D_1 D_2 D_3},
\end{eqnarray*}
where
\begin{eqnarray*}
D_1&=&k^2+i0,\\
D_2&=&(k+p_2)^2+i0,\\
D_3&=&(k+p_2+p_3)^2+i0,
\end{eqnarray*}
and $p_i$ is the four momentum of an $i$'th external particle
(incoming), $k_{\mu}$
the loop momentum, and $\mu_R$ the renormalization energy scale.
An infinitesimal imaginary part ($i0$) is included to obtain 
analyticity of the integral $T^{(3)}_{\mu\cdots\nu}$.
Momentum integration can be performed using Feynman's parameterization,
which combines propagators.
After momentum integration, an ultra-violet pole is subtracted under some 
renormalization scheme. Then space-time dimension is replaced as
$\euv \rightarrow -\eir$ to regulate an infrared pole. 
Finally, the tensor-integral is expressed 
in the following form\cite{graceloop}:
\begin{eqnarray*}
T_{\mu \cdots \nu}^{(3)}&=&\sum_i C_{\mu \cdots \nu}^i 
J_{3}^i(p_1^2,p_2^2,p_3^2;n_x^{(i)},n_y^{(i)}),
\end{eqnarray*}
where
\begin{eqnarray}
J_{3}^i(p_1^2,p_2^2,p_3^2;n_x^{(i)},n_y^{(i)})
&=&\frac{1}{(4 \pi)^2}\frac{\eir \Gamma(-\eir)}{(4 \pi \mu_R^2)^{\eir}}
\int_0^1 dx \int_0^{1-x}  dy
\frac{x^{n_x^{(i)}} y^{n_y^{(i)}}}{D^{1-\eir}}, \nonumber \\
D&=&(p_1 x - p_2 y)^2-\rho x y - p_1^2 x - p_2^2 y -i0, \nonumber \\
\rho&=&p_3^2-(p_1+p_2)^2. 
\end{eqnarray}
The masses of internal particles are assumed to be massless.
The remaining task is to perform the parametric integration of $J_{3}$.
\subsection{Two on-shell, one off-shell external legs}
For the case of two on-shell and one off-shell external particles, we set 
$p_1^2=p_2^2=0,p_3^2\neq0$ without any loss of generality.
The integration can be done in a straightforward way:
\begin{eqnarray}
J_{3}(0,0,p_3^2;n_x,n_y)&=&
\frac{1}{(4 \pi)^2}\frac{\eir \Gamma(-\eir)}{(4 \pi \mu_R^2)^{\eir}}
\int_0^1 dx \int_0^{1-x}  dy
\frac{x^{n_x} y^{n_y}}{(-p_3^2~xy-i0)^{1-\eir}}, \nonumber \\
&=&
\frac{\eir \Gamma(-\eir)}{(4 \pi)^2}
\left(\frac{-\tps}{4 \pi \mu_R^2}\right)^{\eir}
\frac{1}{-p_3^2}
\frac{B(n_x+\eir,n_y+\eir)}{n_x+n_y+2\eir}, \label{v00p}
\end{eqnarray}
where $\tps=p_3^2+i0$, and $B(\cdot,\cdot)$ is a Beta function.
The infrared structure of the tensor-integral can be obtained by
expanding Eq.(\ref{v00p}) with respect to $\eir$. The results of 
expansions under the $\overline{MS}$ scheme are shown in Appendix $B$.
When both $n_x$ and $n_y$ are non-zero, there is no IR-divergence as
\begin{eqnarray*}
J_{3}(0,0,p_3^2;n_x,n_y) 
&\rightarrow&
\frac{1}{(4 \pi)^2 p_3^2 }\frac{(n_x-1)! (n_y-1)!}{(n_x+n_y)!}~~~~~~~~(\eir \rightarrow 0).
\end{eqnarray*}
\subsection{One on-shell, two off-shell external legs}
In the case of one on-shell and two off-shell external particles, we set
$p_1^2=0,p_2^2\neq0,p_3^2\neq0$. The integral $J_{3}$ becomes
\begin{eqnarray}
&~&J_{3}(0,p_2^2,p_3^2;n_x,n_y)\nonumber \\
&=&\frac{1}{(4 \pi)^2}\frac{\eir \Gamma(-\eir)}{(4 \pi \mu_R^2)^{\eir}}
\int_0^1 dx \int_0^{1-x}  dy
\frac{x^{n_x} y^{n_y}}{((p_3^2-p_2^2)xy-p_2^2~y (1-y) -i0)^{1-\eir}}, \nonumber \\
&=&
\frac{\eir \Gamma(-\eir)}{(4 \pi)^2}
\left(\frac{-\tps}{4 \pi \mu_R^2}\right)^{\eir}
\frac{1}{-p_3^2}
\frac{B(n_x+\eir,n_y+\eir)}{n_x+n_y+2\eir} \nonumber \\
&\times&
\hyg\left(1,1-\eir,2+n_x;\frac{p_3^2-p_2^2}{\tps}\right)
\frac{n_x+\eir}{n_x+1},
\nonumber \\
&=&
J_{3}(0,0,p_3^2;n_x,n_y)~{\cal G}_{n_x}\left(\frac{p_3^2-p_2^2}{\tps}\right),
\label{v0pp}
\end{eqnarray}
where
\begin{eqnarray}
{\cal G}_{n}(z)&=&
\frac{n+\eir}{n+1}
\hyg\left(1,1-\eir,2+n;z\right), \label{gfunc}
\end{eqnarray}
and $\hyg(\cdot,\cdot,\cdot;\cdot)$ is the hypergeometric function.
A definition and some properties of the hypergeometric function 
and its numerical 
evaluation can be found in Appendix $A$.
When $p_2^2\rightarrow 0$, the hypergeometric function becomes
\begin{eqnarray*}
\hyg\left(1,1-\eir,2+n_x;z\right)
\rightarrow \frac{n_x+1}{n_x+\eir}~~~(z\rightarrow1).
\end{eqnarray*}
Then, the result Eq.(\ref{v0pp}) agrees with Eq.(\ref{v00p}) 
when $p_2^2\rightarrow 0$.
The infrared structure of the tensor-integral can be obtained by
expanding Eq.(\ref{v0pp}) with respect to $\eir$. The results of 
expansions under the $\overline{MS}$ scheme are shown in Appendix $B$.
\section{Box integral}
\setcounter{equation}{0}
\subsection{Massless one-loop {\it box} integral in n-dimension}
Box integrations can be treated the same as in the vertex case.
The tensor-integral of a massless one-loop vertex with rank $M \leq 4$ 
in space-time
dimensions of $n=4-2\euv$ can be written as
\begin{eqnarray*}
T_{\underbrace{\mu \cdots \nu}_{M}}^{(4)}&=&
\left(\mu_R^2\right)^{\euv} \int  \frac{d^nk}{(2 \pi)^ni}
\frac{k_\mu \cdots k_\nu}{D_1 D_2 D_3 D_4}, 
\end{eqnarray*}
where
\begin{eqnarray*}
D_1&=&k^2+i0,\\
D_2&=&(k+p_1)^2+i0,\\
D_3&=&(k+p_1+p_2)^2+i0,\\
D_4&=&(k+p_1+p_2+p_3)^2+i0.
\end{eqnarray*}
After the same procedure as that 
used in vertex integration, we are left with
following parametric integration:
\begin{eqnarray}
J_{4}(s,t;p_1^2,p_2^2,p_3^2,p_4^2;n_x,n_y,n_z)&=&
\frac{\Gamma(2-\eir)} {(4 \pi)^2
\left(4 \pi \mu_R^2\right)^{\eir}}
\int_0^1 dx~\int_0^{1-x}dy~\int_0^{1-x-y}dz
\frac{x^{n_x}y^{n_y}z^{n_z}}{D^{2-\eir}}, \nonumber \\
\label{box}
\end{eqnarray}
where
\begin{eqnarray*}
D&=&-s~xz-t~y(1-x-y-z)-p_1^2~xy-p_2^2~yz \\
&~&-p_3^2~z(1-x-y-z)-p_4^2~x(1-x-y-z)-i0, \\
s&=&(p_1+p_2)^2, \\
t&=&(p_1+p_4)^2.
\end{eqnarray*}
\subsection{Four on-shell external legs}
When all external particles are on-shell (massless),
$p_1^2=p_2^2=p_3^2=p_4^2=0$, the integral of Eq.(\ref{box}) can be
\begin{eqnarray}
&~&J_4(s,t;0,0,0,0;n_x,n_y,n_z)=\nonumber  \\
&~&
\frac{\Gamma(2-\eir)} {(4 \pi)^2
\left(4 \pi \mu_R^2\right)^{\eir}}
\int_0^1 dx~dy~dz
\frac{x^{n_x}y^{n_y}z^{n_z}}
{\left(
-xzs-y(1-x-y-z)t-i0
\right)^{2-\eir}}. \label{box4onshell}
\end{eqnarray}
After applying the transformation
\begin{eqnarray*}
x&=&r~v, \\
y&=&w~(1-r), \\
z&=&(1-r)~(1-w),
\end{eqnarray*}
the integral becomes
\begin{eqnarray*}
J_4(s,t;0,0,0,0;n_x,n_y,n_z)&=&
\frac{\Gamma(2-\eir)} {(4 \pi)^2
\left(4 \pi \mu_R^2\right)^{\eir}}
\int_0^1 dr~r^{-1+n_x+\eir}(1-r)^{-1+n_y+n_z+\eir} \nonumber \\
&\times&\int_0^1 dv\int_0^1 dw \frac{w^{n_y}(1-w)^{n_z}v^{n_x}}
{\left(-s~v(1-w)-t~(1-v)w-i0\right)^{2-\eir}} \\
&=&
\frac{\Gamma(2-\eir)} {(4 \pi)^2}
\left(4 \pi \mu_R^2\right)^{-\eir}
B(n_x+\eir,n_y+n_z+\eir) \\
&\times&\int_0^1 dv\int_0^1 dw \frac{w^{n_y}(1-w)^{n_z}v^{n_x}}
{(-s~v(1-w)-t~(1-v)w-i0)^{2-\eir}}.
\end{eqnarray*}
The $r$-integral just gives the Beta function.
Then, the $v$-integral can be done as
\begin{eqnarray*}
I_v&\equiv&
\int^1_0 dv \frac{v^{n_x}} 
{\left[\left(-s+s~w+t~w\right)v-t~w-i0\right]^{2-\eir}}  \\
&=&\frac{\left(-\t\right)^{\eir}w^{\eir-2}}{t^2(1+n_x)}
\hyg\left(2-\eir,1+n_x,2+n_x,-\xiwt\right), \label{vint}
\end{eqnarray*}
where
\begin{eqnarray*}
\xiwt&=&\frac{\u}{\t}+\frac{\s}{\t~w}, \\
{\s}&=&s+i0,\\
{\t}&=&t+i0,\\
{\u}&=&u+i0=(p_1+p_3)^2+i0.
\end{eqnarray*}
Here, we use $s+t+u=\sum_{i}p_i^2=0$.

Further integration of the hype-geometric function is not straightforward.
When $a$ (or $b$) in the hypergeometric series $\hyg(a,b,c;z)$ is
a negative integer, the hypergeometric series is truncated at some point,
and becomes a polynomial.
In oder to express our target integrand as a polynomial,
a transformation formula,
\begin{eqnarray*}
&~&\hyg\left(a,b,c;z\right)=\frac{\Gamma(c)\Gamma(c-a-b)}{\Gamma(c-a)
\Gamma(c-b)}z^{-a}\hyg\left(a,a-c+1,a+b-c+1;1-\frac{1}{z}\right) \nonumber \\
&+&\frac{\Gamma(c)\Gamma(a+b-c)}{\Gamma(a)\Gamma(b)}
(1-z)^{c-a-b}z^{a-c}\hyg\left(c-a,1-a,c-a-b+1;1-\frac{1}{z}\right), \label{hgft} \\
&~&(\left|arg~z\right|<\pi,\left|arg~(1-z)\right|<\pi),
\end{eqnarray*}
is used. After this transformation, the hypergeometric function becomes
\begin{eqnarray}
&~&\hyg\left(2-\eir,1+n_x,2+n_x,-\xiwt\right)= 
(n_x+1)! \Gamma(\eir-1)
\nonumber \\
&\times& \biggl[\frac{1}{\Gamma(n_x+\eir)}\left(-\xiwt\right)^{-1-n_x} 
+\frac{(1+\xiwt)^{\eir-1}} \xiwt\sum_{l=0}^{n_x}
\left(1+\frac{1}{\xiwt}\right)^l
\frac{(-1)^l}{\Gamma(l+\eir) (n_x-l)!}\biggr]. 
\nonumber \\ \label{hygex}
\end{eqnarray}
Though the hypergeometric
function in l.h.s.~of Eq.(\ref{hygex}) is finite when $n_x\geq1$ and
$\eir \rightarrow 0$,
the gamma function in r.h.s.~has a $1/\eir$ pole.
We have confirmed that the terms in brackets on the r.h.s.~of Eq.(\ref{hygex})
start ${\cal O}(\eir)$ when $n_x\geq1$. Thus, there is no $1/\eir$ pole,
as expected. 

Then, the remaining $w$-integral in $J_{4}$ becomes
\begin{eqnarray*}
&~&J_4(s,t;0,0,0,0;n_x,n_y,n_z)= \\
&~&\frac{-1} {(4 \pi~t)^2 }
\left(\frac{-\t}{4 \pi \mu_R^2}\right)^{\eir}
B(n_x+\eir,n_y+n_z+\eir) n_x! \Gamma(\eir) \Gamma(1-\eir) \\
&\times&\int_0^1dw~w^{n_y-2+\eir}(1-w)^{n_z} 
\biggl[\frac{1}{\Gamma(n_x+\eir)}\left(-\xiwt\right)^{-1-n_x}\\
&+&\frac{\left(1+\xiwt\right)^{\eir-1}}{\xiwt}
\sum_{l=0}^{n_x}\left(1+\frac{1}{\xiwt}\right)^l 
\frac{(-1)^l}{\Gamma(l+\eir)(n_x-l)!}
\biggr].
\end{eqnarray*}
The $w$-integral can be solved in a term-by-term way for each power of $l$.
The final form of $J_{4}$ is obtained to be
\begin{eqnarray}
&~&J_4(s,t;0,0,0,0;n_x,n_y,n_z)
=\frac{1}{(4\pi)^2s~t}B(n_x+\eir,n_y+n_z+\eir)n_x! \Gamma(\eir) \Gamma(1-\eir)
\nonumber \\
&\times&\biggl[
\left(\frac{-\t}{4 \pi \mu_R^2}\right)^{\eir}
\left(\frac{-t}{s}\right)^{n_x} 
\frac{B(1+n_z,n_x+n_y+\eir)}{\Gamma(n_x+\eir)}
\nonumber \\
&\times&\hyg\left(1+n_x,n_x+n_y+\eir,1+n_x+n_y+n_z+\eir,-\frac{\u}{\s}\right)
\nonumber \\
&+&\left(\frac{-\s}{4 \pi \mu_R^2}\right)^{\eir}
\sum_{l=0}^{n_x} \left(\frac{-s}{t}\right)^l 
\frac{(-1)^l}{\Gamma(l+\eir)(n_x-l)!}
B(1+n_y,l+n_z+\eir)
\nonumber \\
&\times&\hyg\left(1+l,l+n_z+\eir,1+l+n_y+n_z+\eir,-\frac{\u}{\bar t}\right)
\biggr], \label{b0-nnn}
\end{eqnarray}
where ${\bar t}=t-i0$.

When the numerator of the integrand is unity, $n_x=n_y=n_z=0$,
the result of Eq.(\ref{b0-nnn}) is reduced to
\begin{eqnarray}
&~&J_{4}(s,t;0,0,0,0;0,0,0)=
\frac{1} {(4 \pi)^2 s~t}
\frac{B(\eir,\eir)\Gamma(1-\eir)}{\eir}
\nonumber \\
&\times&\left[\left(\frac{-\s}{4 \pi \mu_R^2}\right)^{\eir}
\hyg\left(1,\eir,1+\eir,-\frac{\u}{\bar t}\right)
+\left(\frac{-\t}{4 \pi \mu_R^2}\right)^{\eir}
\hyg\left(1,\eir,1+\eir,-\frac{\u}{\s}\right)\right]. \nonumber \\
\label{b0-000}
\end{eqnarray}
We have checked that this result agrees with the precedence result obtained
by Duplan$\breve{z}$i$\acute{c}$ and Ni$\breve{z}$i$\acute{c}$\cite{d&n}
in both physical and unphysical regions of kinematical 
variables, $s$ and $t$.

When $n_x=0$, the result of Eq.(\ref{b0-nnn}) has a shorter expression
without a fake pole, as
\begin{eqnarray}
&~&J_{4}(s,t;0,0,0,0;0,n_y,n_z)
=\frac{1}{(4\pi)^2s~t}B(\eir,n_y+n_z+\eir)\Gamma(1-\eir)
\nonumber \\
&\times&\biggl[
\left(\frac{-\t}{4 \pi \mu_R^2}\right)^{\eir}
B(1+n_z,n_y+\eir)
\hyg\left(1,n_y+\eir,1+n_y+n_z+\eir,-\frac{\u}{\s}\right)
\nonumber \\
&+&~\left(\frac{-\s}{4 \pi \mu_R^2}\right)^{\eir}
B(1+n_y,n_z+\eir)
\hyg\left(1,n_z+\eir,1+n_y+n_z+\eir,-\frac{\u}{\bar t}\right)
\biggr]. \label{b0-0nn}
\end{eqnarray}
The infrared behavior of the loop integral can be obtained by
expanding this formula with respect to $\eir$, as shown in Appendix $B$.

In some cases with $n_x\neq0$, we can avoid the fake pole by
using the symmetry of the integrand.
The basic integrand, Eq.(\ref{box4onshell}), is symmetric under the
exchange $x(n_x) \leftrightarrow z(n_z)$. Then, the result with
$n_x\neq0$ and $n_z=0$ can be easily obtained as
\begin{eqnarray*}
J_{4}(s,t;0,0,0,0;n_x,n_y,0)
&=&\{J_4(s,t,0,0,0,0;0,n_y,n_z),n_z \rightarrow n_x\}
\end{eqnarray*}
for any values of $n_y$.

In the case of $n_x\neq0,n_z\neq0$, the integration has no
IR-divergence with any value of $n_y\geq0$.
However we cannot simply set $\eir$ to be zero, because there is the fake pole.
After expanding Eq.(\ref{b0-nnn}) with respect to $\eir$, we confirmed that
the $\eir$ pole was canceled out. An explicit form of the Tyler expansion of
the hypergeometric functions can be found in Appendix $A$.
\subsection{1 off-shell and 3 on-shell external legs}
When one of four external particles is off-shell, we can set it to be
$p_1$ without any loss of generality,
$p_1^2\neq0, p_2^2=p_3^2=p_4^2=0$. 
Then, the integral of Eq.(\ref{box}) can be
\begin{eqnarray}
&~&J_4(s,t;p_1^2,0,0,0;n_x,n_y,n_z)= \nonumber \\
&~&\frac{\Gamma(2-\eir)} {(4 \pi)^2
\left(4 \pi \mu_R^2\right)^{\eir}}
\int_0^1 dx~\int_0^{1-x}dy~\int_0^{1-x-y}dz
\frac{x^{n_x}y^{n_y}z^{n_z}}
{\left(-xzs-y(1-x-y-z)t-p_1^2xy-i0 \right)^{2-\eir}}. \nonumber \\
\label{box2}
\end{eqnarray}
After applying the same transformation as for the four-on-shell case,
the integral becomes
\begin{eqnarray*}
J_4(s,t;p_1^2,0,0,0;n_x,n_y,n_z)
&=&
\frac{\Gamma(2-\eir)} {(4 \pi)^2
\left(4 \pi \mu_R^2\right)^{\eir}}
B(n_x+\eir,n_y+n_z+\eir) \\
&\times& \int_0^1 dv\int_0^1 dw \frac{w^{n_y}(1-w)^{n_z}v^{n_x}}
{(-s~v(1-w)-t~(1-v)w-p_1^2~vw-i0)^{2-\eir}}.
\end{eqnarray*}
The $v$-integral can be done as
\begin{eqnarray*}
I_v
&=&
\int^1_0 dv \frac{v^{n_x}} 
{\left[\left(-s+s~w+t~w-p_1^2~w\right)v-t~w-i0\right]^{2-\eir}}  \\
&=&\frac{\left(-\t\right)^{\eir}w^{\eir-2}}{t^2(1+n_x)}
\hyg\left(2-\eir,1+n_x,2+n_x,-\xiwt\right), 
\end{eqnarray*}
where
\begin{eqnarray*}
\xiwt&=&\frac{\u}{\t}+\frac{\s}{\t~w}.
\end{eqnarray*}
Here, we use $s+t+u=p_1^2$.
This result is the same as that in the four-on-shell case, except that
$u=-s-t+p_1^2$ instead of $u=-s-t$.
After making the same transformation as that for the four-on-shell case, 
the remaining $w$-integral can be expressed as
\begin{eqnarray}
&~&J_4(s,t;p_1^2,0,0,0;n_x,n_y,n_z) =
\frac{1}{(4\pi)^2s~t}B(n_x+\eir,n_y+n_z+\eir)n_x! \Gamma(\eir) \Gamma(1-\eir)
\nonumber \\
&\times&\biggl[
\left(\frac{-\t}{4 \pi \mu_R^2}\right)^{\eir}
\left(\frac{-t}{s}\right)^{n_x} 
\frac{1}{\Gamma(n_x+\eir)}\Io 
+\left(\frac{-\s}{4 \pi \mu_R^2}\right)^{\eir} 
\sum_{l=0}^{n_x} 
\frac{(-1)^l}{\Gamma(l+\eir)(n_x-l)!}\It
\biggr], 
\nonumber \\ \label{b1-nnn}
\end{eqnarray}
where
\begin{eqnarray*}
\Io&=&\int_0^1 dw~w^{n_x+n_y-1+\eir}\left(1-w\right)^{n_z} 
\left(1+\frac{\u}{\s}w\right)^{-1-n_x}, \\
\It&=&
\frac{-t}{s}
\int_0^1 dw~w^{n_y}\left(1-w\right)^{n_z} 
\left(1+\frac{\u}{\s}w\right)^{-1-l}\left(1+\frac{\t+\u}{\s}w\right)^{l-1+\eir}.
\end{eqnarray*}
The first integration, $\cal{I}^{\rm (1)}$, can be done as
\begin{eqnarray}
\Io&=&B(1+n_z,n_x+n_y+\eir)
\hyg\left(1+n_x,n_x+n_y+\eir,1+n_x +n_y+n_z+\eir,-\frac{\u}{\s}\right).\nonumber \\
\label{b1-nnn-I1}
\end{eqnarray}
For the second integration, $\It$, we used a following binomial expansion:
\begin{eqnarray*}
\left(1-w\right)^{n_z}w^{n_y}&=&\sum_{k_1=0}^{n_z}~_{n_z}C_{k_1}~
\left(-1\right)^{k_1}
w^{n_y+k_1}, \\
w^{n_y+k_1}&=&\left(\frac{s}{t+u}\right)^{n_y+k_1}
\left(
\left(1+\frac{\t+\u}{\s}w\right)-1
\right)^{n_y+k_1} \\ 
&=&\left(\frac{s}{p_1^2-s}\right)^{n_y+k_1}
\sum_{k_2=0}^{n_y+k_1} ~_{n_y+k_1}C_{k_2}~(-1)^{n_y+k_1-k_2}~
\left(1+\frac{\t+\u}{\s} w \right)^{k_2},
\end{eqnarray*}
where $_{m}C_{n}$ is combinatorial defined as
\begin{eqnarray}
~_{m}C_{n}&\equiv&\frac{m!}{n! (m-n)!}.\label{mCn}
\end{eqnarray}

Then, the second integration, $\It$, can be done as
\begin{eqnarray}
\It&=&
\sum_{k_1=0}^{n_z}~_{n_z}C_{k_1}~
\left(\frac{s}{p_1^2-s}\right)^{n_y+k_1}
\sum_{k_2=0}^{n_y+k_1} ~_{n_y+k_1}C_{k_2}~ (-1)^{n_y+k_2}
\left(\frac{-t}{s}\right)\nonumber \\
&\times&\int_0^1 dw~\left(1+\frac{\u}{\s}w\right)^{-(l+1)}
\left(1+\frac{\t+\u}{\s}w\right)^{k_2+l-1+\eir}
\nonumber \\
&=&\sum_{k_1=0}^{n_z}
\sum_{k_2=0}^{n_y+k_1} 
~_{n_z}C_{k_1}~
~_{n_y+k_1}C_{k_2}~ 
(-1)^{k_1+k_2}~
\left(\frac{s}{s-p_1^2}\right)^{n_y+k_1}
\frac{1}{l+k_2+\eir}
\left(1+\frac{u}{t}\right)^{l}\nonumber \\
&\times&\biggl[\hyg\left(1+l,l+k_2+\eir,1+l+k_2+\eir,-\frac{\u}{\bar t}\right)
\nonumber \\
&-&\left(\frac{\tpo}{\s}\right)^{l+k_2+\eir}
\hyg\left(1+l,l+k_2+\eir,1+l+k_2+\eir,-\frac{\u \tpo}{{\bar t}\s}\right)
\biggr], \label{b1-nnn-I2}
\end{eqnarray}
where $\tpo=p_1^2+i0$. 
Here, the integral $J_{4}(s,t;p_1^2,0,0,0;n_x,n_y,n_z)$ 
can be successfully expressed by 
a finite number of hypergeometric functions.
The infrared structure of the tensor-integral can be obtained by
expanding Eq.(\ref{b1-nnn}) with respect to $\eir$. 
The results of 
expansions under the $\overline{MS}$ scheme are given in Appendix $B$.

When the numerator of the integrand is unity, $n_x=n_y=n_z=0$,
the second integration Eq.(\ref{b1-nnn-I2}) is reduced to
\begin{eqnarray}
\cal{I}^{\rm (2)}_0&=&
\frac{-t}{s}
\int_0^1 dw~
\left(1+\frac{\u}{\s}w\right)^{-1}\left(1+\frac{\t+\u}{\s}w\right)^{\eir-1}
\nonumber \\
&=&\frac{1}{\eir}\left[
\hyg\left(1,\eir,1+\eir,-\frac{\u}{\bar t}\right)
-\left(\frac{{\tpo}}{\s}\right)^{\eir}
\hyg\left(1,\eir,1+\eir,-\frac{\u \tpo}{{\bar t}\s}\right)
\right].  \nonumber \\
\end{eqnarray}
Then, the Eq.(\ref{b1-nnn}) can be written as
\begin{eqnarray}
&~&J_4(s,t;p_1^2,0,0,0;0,0,0)=\frac{1}{(4\pi)^2s~t}
\frac{B(\eir,\eir)\Gamma(1-\eir)}{\eir}
\nonumber \\
&\times&\biggl[\left(\frac{-\s}{4 \pi \mu_R^2}\right)^{\eir}
\hyg\left(1,\eir,1+\eir,-\frac{\u}{\bar t}\right)
+\left(\frac{-\t}{4 \pi \mu_R^2}\right)^{\eir}
\hyg\left(1,\eir,1+\eir,-\frac{\u}{\s}\right)\nonumber \\
&-&\left(\frac{-\tpo}{4 \pi \mu_R^2}\right)^{\eir}
\hyg\left(1,\eir,1+\eir,-\frac{\u \tpo}{{\bar t}\s}\right)
\biggr], \nonumber \\
&=&J_4(s,t;0,0,0,0;0,0,0)\nonumber \\
&-&\frac{1}{(4\pi)^2s~t}
\frac{B(\eir,\eir)\Gamma(1-\eir)}{\eir}
\left(\frac{-\tpo}{4 \pi \mu_R^2}\right)^{\eir}
\hyg\left(1,\eir,1+\eir,-\frac{\u \tpo}{{\bar t}\s}\right)
\biggr]. 
\end{eqnarray}
We have again checked that this result agrees with the 
precedence result obtained
by Duplan$\breve{z}$i$\acute{c}$ and Ni$\breve{z}$i$\acute{c}$\cite{d&n}
in both the physical and unphysical regions of the kinematical variables.

When $n_x=0$, the result of Eq.(\ref{b1-nnn}) does not have a fake pole.
The result is obtained to be
\begin{eqnarray}
&~&J_4(s,t;p_1^2,0,0,0;0,n_y,n_z)
=\frac{1}{(4\pi)^2s~t}B(\eir,n_y+n_z+\eir) \Gamma(1-\eir)
\nonumber \\
&\times&\left[
\left(\frac{-\t}{4 \pi \mu_R^2}\right)^{\eir}
\Io
+\left(\frac{-\s}{4 \pi \mu_R^2}\right)^{\eir}
{\cal I}^{\rm (2)}_0
\right], \label{b1-0nn}
\end{eqnarray}
where
\begin{eqnarray}
\Io&=&B(1+n_z,n_y+\eir)
\hyg\left(1,n_y+\eir,1+n_y+n_z+\eir,-\frac{\u}{\s}\right).
\label{b1-0nn-I12}
\\
{\cal I}^{\rm (2)}_0
&=&\sum_{k_1=0}^{n_z}
\sum_{k_2=0}^{n_y+k_1} 
~_{n_z}C_{k_1}~
~_{n_y+k_1}C_{k_2}~ 
(-1)^{k_1+k_2}~
\left(\frac{s}{s-p_1^2}\right)^{n_y+k_1}
\frac{1}{k_2+\eir}
\nonumber \\
&\times&\biggl[\hyg\left(1,k_2+\eir,1+k_2+\eir,-\frac{\u}{\bar t}\right)
-\left(\frac{\tpo}{\s}\right)^{\eir}
\hyg\left(1,k_2+\eir,1+k_2+\eir,-\frac{\u \tpo}{{\bar t}\s}\right)
\biggr]. 
\nonumber \\ 
\end{eqnarray}

In the case of $n_x\neq0,n_z\neq0$, the integration has no
IR-divergence when using any value of $n_y\geq0$.
However, we cannot simply set $\eir$ to be zero again, 
because there is the fake pole.
After expanding Eq.(\ref{b1-nnn}) with respect to $\eir$, we confirmed 
numerically that the $\eir$ pole was canceled out. 
The results of 
expansions under the $\overline{MS}$ scheme are shown in Appendix $B$.
\section{Numerical check of the results}
The results of the vertex tensor-integral are rather trivial. However,
those of the box tensor-integral are very complicated and highly 
non-trivial. We need some cross-checking of our results compared with other
independent calculations.
For the scalar integral of the one or two off-shell box integral,
we can check our results numerically with those of the precedence calculation 
done by Duplan$\breve{z}$i$\acute{c}$ and Ni$\breve{z}$i$\acute{c}$\cite{d&n}.
In both in the physical and unphysical regions of the kinematical 
variables, $s$ and $t$, it was confirmed that the results given in this report
agree completely with those in ref.\cite{d&n}.

The basic ingredients of a numerical calculation 
of the general case of the tensor integral
are given in Appendix $A$. Those formulas are 
Eqs.(\ref{hyg5})$\sim$(\ref{hyg4}) 
to evaluate the hypergeometric function
and Eqs.(\ref{ghyg2})$\sim$(\ref{ghyg7})
to evaluate the generalized hypergeometric functions.

At first, the results of a numerical evaluation of 
the hypergeometric function
of the type Eq.(\ref{hyg3}) are compared with those obtained from 
$Mathematica$\cite{mathematica} at several values of $l,~m,~n$ and $z$ at
random. We confirmed 
that both results agree very well with each other to more than ten digits.
For the generalized hypergeometric function,
our recursion relation formulas, Eqs.(\ref{ghyg2})$\sim$(\ref{ghyg7}),
were checked by comparing 
those of numerical integration of Eq.(\ref{ghyg1}) using
the numerical contour-integral (NCI) method\cite{nci} developed by the author. 
The function ${\tilde F}^{(n)}_{l,m}(z)$, where $n=1,2$ and $l=2,~m=3$,
given in Eq.(\ref{ghyg1}), was numerically evaluated at several values
of $z$, as shown in Tables~1 and 2, which was compared with the NCI results.
Both results gave very good agreement to about ten digits, as shown in
tables.
The imaginary part of the result must be zero, except $z=2000+0i$ case 
in the table. It was also numerically confirmed very precisely.
\setcounter{equation}{0} 
\begin{center}
\begin{tabular}{|c|c||c||c|}
  \hline
$z$ & real/imag.&analytic& NCI \\ \hline \hline
$~2000+10^{-15}i$~
& real &$~3.167042847\times10^{-7}$&$~3.167042848\times10^{-7}$ \\
& imag.&$-1.308996938\times10^{-7}$&$-1.308996939\times10^{-7}$ \\ \hline
$-2000+10^{-15}i$~
& real &$~3.167042847\times10^{-7}$&$~3.167042848\times10^{-7}$ \\
& imag.&   ${\cal O}(10^{-24})$    & ${\cal O}(10^{-18})$       \\ \hline
$ ~0.2+10^{-15}i$~
& real &$~3.723185361\times10^{-1}$&$~3.723185362\times10^{-1}$ \\
& imag.&   ${\cal O}(10^{-16})$    & ${\cal O}(10^{-12})$       \\ \hline
$ -0.2+10^{-15}i$~
& real &$~1.809694496\times10^{-1}$&$~1.809694496\times10^{-1}$ \\
& imag.&   ${\cal O}(10^{-16})$    & ${\cal O}(10^{-12})$       \\ \hline
\end{tabular}
\end{center}
{\bf Table~1}
{\footnotesize Numerical comparison of ${\tilde F}^{(1)}_{2,3}(z)$
between analytic formulas given in Eqs.(\ref{ghyg3})$\sim$(\ref{ghyg4})
and the numerical contour integral of Eq.(\ref{ghyg2}).}\\
\begin{center}
\begin{tabular}{|c|c||c||c|}
  \hline
$z$ & real/imag.&analytic& NCI \\ \hline \hline
$~2000+10^{-15}i$~
& real &$~1.024874615\times10^{-6}$&$~1.024874616\times10^{-6}$ \\
& imag.&$-9.949558053\times10^{-7}$&$-9.949558053\times10^{-7}$ \\ \hline
$-2000+10^{-15}i$~
& real &$~1.230491374\times10^{-6}$&$~1.230491374\times10^{-6}$ \\
& imag.&   ${\cal O}(10^{-24})$    & ${\cal O}(10^{-17})$       \\ \hline
$ ~0.2+10^{-15}i$~
& real &$~1.624874690\times10^{-1}$&$~1.624874690\times10^{-1}$ \\
& imag.&   ${\cal O}(10^{-15})$    & ${\cal O}(10^{-13})$       \\ \hline
$ -0.2+10^{-15}i$~
& real &$~1.005561296\times10^{-1}$&$~1.005561296\times10^{-1}$ \\
& imag.&   ${\cal O}(10^{-15})$    & ${\cal O}(10^{-12})$       \\ \hline
\end{tabular}
\end{center}
{\bf Table~2}
{\footnotesize Numerical comparison of ${\tilde F}^{(2)}_{2,3}(z)$
between analytic formulas given in Eqs.(\ref{ghyg6})$\sim$(\ref{ghyg7})
and the numerical contour integral of Eq.(\ref{ghyg6}).}\\
\vskip 0.5cm
\begin{center}
\begin{tabular}{|ccc|c||c||c|}
  \hline
$n_x$&$n_y$&$n_z$& real/imag.&analytic& NCI \\ \hline \hline
1&2&3
& real &$-2.15298\times10^{-9}$&$-2,15297\times10^{-9}$ \\
~&~&~
& imag.&$-2.78647\times10^{-9}$&$-2.78650\times10^{-9}$ \\ \hline
2&0&2
& real &$~~~9.74570\times10^{-9}$&$~~~9.74572\times10^{-9}$ \\
~&~&~
& imag.&$-3.22229\times10^{-8}$&$-3.22230\times10^{-8}$ \\ \hline
\end{tabular}
\end{center}
{\bf Table~3}
{\footnotesize Numerical comparison of $J_4(s,t;0,0,0,0;n_x,n_y,n_z)$
between analytic formulas given in Eq.(\ref{res1})
and the numerical contour integral. Here, we set kinematical variables in
the physical region at $s=123$, $t=-200$ and $\mu_R=1$.}
\begin{center}
\begin{tabular}{|ccc|c||c||c|}
  \hline
$n_x$&$n_y$&$n_z$& real/imag.&analytic& NCI \\ \hline \hline
1&2&3
& real &$-7.88683\times10^{-10}$&$-7.88689\times10^{-10}$ \\
~&~&~
& imag.&$-1.95176\times10^{-9}$&$-1.95176\times10^{-9}$ \\ \hline
2&0&2
& real &$~~~1.48133\times10^{-8}$&$~~~1.48133\times10^{-8}$ \\
~&~&~
& imag.&$-2.04318\times10^{-8}$&$-2.04318\times10^{-8}$ \\ \hline
\end{tabular}
\end{center}
{\bf Table~4}
{\footnotesize Numerical comparison of $J_4(s,t;p_1^2,0,0,0;n_x,n_y,n_z)$
between analytic formulas given in Eq.(\ref{res2})
and the numerical contour integral. Here, we set the kinematical variables in
the physical region $s=123$, $t=-200$ $p_1^2=80$ and $\mu_R=1$.}
\vskip 0.5cm
The numerical results of the box tensor-integral 
with zero and one off-shell external
legs obtained using Eqs.({\ref{res1}) and ({\ref{res2}) are also
compared with those using the NCI method, integrating 
Eqs.(\ref{box}) and (\ref{box2}) directly. Both results show very good
agreement within about five digits, as shown in Tables~3 and 4. 
\section{Conclusions}
\setcounter{equation}{0}
The general formulas of the 3- and 4-point tensor-integral were
obtained directly without any reduction method to the scalar integrals.
The IR behavior of the tensor integrals was clearly shown by
expanding the results with respect to the extra space-time dimension
due to the dimensional regularization. All results 
were expressed by very short formulas in  
a suitable manner for a numerical calculation.
The results of the scalar-integral were compared with the precedence results,
and showed complete agreement in both physical and unphysical regions
of the kinematical variables.
For the IR finite case, the analytic results were compared with
the numerical contour integration, and gave a consistent result
within the numerical integration error.
\vskip 1cm
\noindent
{\large \bf Acknowledgments}\\
\vskip 0.2cm
The author would like to thank Drs.~J.~Fujimoto, T.~Ishikawa, T.~Kaneko,
S.~Odaka and Y.~Shimizu
for continuous discussions
on this subject and their useful suggestions.
We are grateful to
Prof.~J.~Vermaseren for his kind help to give us a program
to calculate a tri-logarithmic function.

This work was supported in part by the Ministry of Education, Science
and Culture under the Grant-in-Aid No. 11206203 and 14340081.
This work was also supported in part by the ``International
Research Group'' on ``Automatic Computational Particle Physics''
(IRG ACPP) funded by CERN/Universities in France,
KEK/MEXT in Japan and MSU/RAS/MFBR in Russia.

\newpage
\appendix{\bf \Huge Appendix}
\section{Numerical calculation of the hypergeometric function}
In this Appendix, the basic properties of the hypergeometric 
function\cite{hgf} and
their numerical evaluation are summarized. The $1/\eir$ expansion of the
hypergeometric functions appearing in the tensor-integrals and their
numerical evaluation are also shown.
\subsection{Hypergeometric function}
\setcounter{equation}{0}
The Gauss-series representation of the hypergeometric function is
\begin{eqnarray}
\hyg(a,b,c;z)&=& \hyg(b,a,c;z) \nonumber \\
&=&\sum_{k=0}^{\infty}\frac{(a)_k(b)_k}{(c)_k}\frac{z^k}{k!}, \label{hyg1}
\end{eqnarray}
where $(\cdot)_k$ is Pochhammer's symbol defined as
\begin{eqnarray}
(a)_k&=&\frac{\Gamma(a+k)}{\Gamma(a)}.\nonumber
\end{eqnarray}
The Euler integral representation is
\begin{eqnarray}
\hyg(a,b,c;z)&=& \frac{\Gamma(c)}{\Gamma(b)\Gamma(c-b)}
\int_0^1~\tau^{b-1}(1-\tau)^{c-b-1}(1-z\tau)^{-a}~d\tau, \label{hyg2}\\
&~&({\cal R}c>{\cal R}b>0).\nonumber
\end{eqnarray}

When $a$ is a negative integer, such as $a=-m$, the Gauss series is truncated
at $k=m$, and becomes a polynomial,
\begin{eqnarray}
\hyg(-m,b,c;z)&=&
\sum_{k=0}^m \frac{(-m)_k(b)_k}{(c)_k}\frac{z^k}{k!}
\nonumber \\
&=& \sum_{k=0}^m \frac{(b)_k}{(c)_k}~_mC_k(-z)^k,
\label{hyg5}
\end{eqnarray}
where $_{m}C_{k}$ is combinatorial defined in Eq.(\ref{mCn}).
For numerical evaluations of tensor-integrals, 
the following type of the hypergeometric
function might be numerically calculated as
\begin{eqnarray}
&~&\hyg(l,m+1,n+m+2;z)\nonumber \\
&=&\frac{1}{B(m+1,n+1)}
\int_0^1\tau^m (1-\tau)^n (1-z\tau)^{-l} d\tau \nonumber \\
&=&\sum_{k_1=0}^n \sum_{k_2=0}^{m+k_1}(-1)^{k_1+k_2}
\frac{_nC_{k_1}~_{m+k_1}C_{k_2}}{B(m+1,n+1)}\frac{1}{z^{m+k_1}}
\int_0^1(1-z\tau)^{-l+k_2}d\tau,\label{hyg3}
\end{eqnarray}
where $l,m,n$ are positive integers.
Here, integration can be performed as
\begin{equation}
\int_0^1(1-z\tau)^{-l+k_2}d\tau=
\begin{cases}
-\frac{\ln(1-z)}{z}  & k_2-l+1=0, \\ 
\frac{1}{k_2-l+1}\frac{(1-z)^{k_2-l+1}-1}{-z} & k_2-l+1\neq0.\label{hyg4}
\end{cases}
\end{equation}
This formulas can be used for a numerical evaluation of hypergeometric
functions of this type.

\subsection{Generalized hypergeometric function}
For a Laurant expansion of the hypergeometric function with respect to
$\eir$, the following generalized hypergeometric function is necessary:
\begin{eqnarray}
&~&_3F_2(\{a_1,a_2,a_3\},\{b_1,b_2\};z)\nonumber \\
&=&
\sum_{n=0}^{\infty}\frac{(a_1)_n(a_2)_n(a_3)_n}{(b_1)_n(b_2)_n}
\frac{z^n}{n!}\\
&=&\frac{\Gamma(b_1)\Gamma(b_2)}
{\Gamma(a_1)\Gamma(b_1-a_1)\Gamma(a_2)\Gamma(b_2-a_2)}\nonumber \\
&\times&\int_0^1 d\tau~\int_{\tau}^1dv~v^{a_1-b_2}(1-v)^{b_1-a_1-1}\tau^{a_2-1}
(v-\tau)^{b_2-a_2-1}(1-z \tau)^{-a_3},
\end{eqnarray}
and
\begin{eqnarray}
_4F_3(\{a_1,a_2,a_3,a_4\},\{b_1,b_2,b_3\};z)
&=&
\sum_{n=0}^{\infty}\frac{(a_1)_n(a_2)_n(a_3)_n(a_4)_n}{(b_1)_n(b_2)_n(b_3)_n}
\frac{z^n}{n!}.
\end{eqnarray}
%
%
Those generalized hypergeometric functions appear in a
following integral:
\begin{eqnarray}
{\tilde F}^{(n)}_{j_1,j_2}(z)&\equiv&\frac{(-1)^n}{n!}
\int_0^1d\tau~\tau^{j_1-1}(1-z \tau)^{-(j_2+1)} \ln^n{\tau},
\label{ghyg1}
\end{eqnarray}
where $j_1$ is a positive integer and $j_2$ is an integer  
(it can be negative).

When $n=1$, the integral becomes
\begin{eqnarray}
{\tilde F}^{(1)}_{j_1,j_2}(z)&=&-
\int_0^1d\tau~\tau^{j_1-1}(1-z \tau)^{-(j_2+1)} \ln{\tau}, 
\label{ghyg2} \\
&=&
\frac{_3F_2(\{j_1,j_1,j_2+1\},\{j_1+1,j_1+1\};z)}{j_1^2},\\
&=&\sum_{k=0}^{\infty}\frac{z^k}{(j_1+k)^2(j_2+k+1) B(j_2+1,k+1)}.
\end{eqnarray}
When $j_2$ is a negative integer, such as $j_2=-j\leq-1$, this function can be
express by a polynomial,
\begin{eqnarray}
{\tilde F}^{(1)}_{j_1,-j}(z)
&=&\sum_{k=0}^{j-1}\frac{_{j-1}C_k~(-z)^k}{(j_1+k)^2}.
\label{ghyg3}
\end{eqnarray}
When $J_1=1$ and $j_2=0,1$, the integral can be performed easily as
\begin{eqnarray}
{\tilde F}^{(1)}_{1,0}(z)&=&\frac{{\rm Li}_2(z)}{z}, \\
{\tilde F}^{(1)}_{1,1}(z)&=&-\frac{\ln(1-z)}{z}.
\end{eqnarray}
When $j_1=1$ and $j_2\geq1$, we can use the following recursion relation:
\begin{eqnarray}
{\tilde F}^{(1)}_{1,j_2+1}(z)&=&\frac{j_2}{j_2+1}{\tilde F}^{(1)}_{1,j_2}
+\frac{(1-z)^{-j_2}-1}{j_2(j_2+1)z}.
\end{eqnarray}
For the general case, the function ${\tilde F}^{(1)}_{j_1,j_2}(z)$ 
can be obtained using
${\tilde F}^{(1)}_{1,\cdot}$,
\begin{eqnarray}
{\tilde F}^{(1)}_{j_1,j_2}(z)&=&\frac{1}{z^{j_1-1}}\sum_{k=0}^{j_1-1}
(-1)^k~_{j_1-1}C_k~{\tilde F}^{(1)}_{1,j_2-k}(z).
\label{ghyg4}
\end{eqnarray}

When $n=2$, the integral becomes
\begin{eqnarray}
{\tilde F}^{(2)}_{j_1,j_2}(z)&=&\frac{1}{2}
\int_0^1d\tau~\tau^{j_1-1}(1-z \tau)^{-(j_2+1)} \ln^2{\tau}
\label{ghyg5}
\\
&=&
\frac{_4F_3(\{j_1,j_1,j_1,j_2+1\},\{j_1+1,j_1+1,j_1+1\};z)}{j_1^3}\\
&=&\sum_{k=0}^{\infty}\frac{z^k}{(j_1+k)^3(j_2+k+1) B(j_2+1,k+1)}.
\end{eqnarray}
When $j_2=-j\leq-1$, it is also represented by a polynomial,
\begin{eqnarray}
{\tilde F}^{(2)}_{j_1,-j}(z)
&=&\sum_{k=0}^{j-1}\frac{_{j-1}C_k~(-z)^k}{(j_1+k)^3}.
\label{ghyg6}
\end{eqnarray}
When $J_1=1$ and $j_2=0,1$, the integral can be performed easily as follows:
\begin{eqnarray}
{\tilde F}^{(2)}_{1,0}(z)&=&\frac{{\rm Li}_3(z)}{z},\\
{\tilde F}^{(2)}_{1,1}(z)&=&\frac{{\rm Li}_2(z)}{z}.
\end{eqnarray}
When $j_1=1$ and $j_2\geq1$, we can use the following recursion relation:
\begin{eqnarray}
(j_2+1){\tilde F}^{(2)}_{1,j_2+1}(z)-j_2{\tilde F}^{(2)}_{1,j_2}(z)
-{\tilde F}^{(1)}_{1,j_2}(z)&=&0.
\end{eqnarray}
For the general case, the function ${\tilde F}^{(2)}_{j_1,j_2}(z)$ can be
obtained using ${\tilde F}^{(2)}_{1,\cdot}(z)$,
\begin{eqnarray}
{\tilde F}^{(2)}_{j_1,j_2}(z)&=&\frac{1}{z^{j_1-1}}\sum_{k=0}^{j_1-1}
(-1)^k~_{j_1-1}C_k~{\tilde F}^{(2)}_{1,j_2-k}(z).
\label{ghyg7}
\end{eqnarray}
\subsection{$1/\eir$ expansion of a hypergeometric function}
In the general form of the tensor integrals, following type of
integral appears:
\begin{eqnarray}
{\cal I}_{l,m,n}&\equiv&
\int_0^1 \tau^{l+n-1+\eir}(1-\tau)^m(1-z \tau)^{-(l+1)}
d\tau,\nonumber \\&=&
B(1+m,l+n+\eir)\hyg\left(1+l,l+n+\eir,1+l+n+m+\eir,z\right).
\end{eqnarray}
In order to show the IR structure of the function, a series expansion
of the function ${\cal I}_{l,m,n}$ with respect to $\eir$ might be considered,
\begin{eqnarray}
{\cal I}_{l,m,n}&=&\sum_{j=-1}^{\infty}{\cal F}^{(j)}_{l,m,n}(z)\eir^{j},
\label{hyg7}
\end{eqnarray}
where $l,m,n$ are non-negative integers. When $l=m=n=0$, the Laurant expansion
of the function can be
\begin{eqnarray}
{\cal I}_{0,0,0}&=&\int_0^1 \frac{(1-z\tau)^{-1}}{\tau^{1-\eir}}d\tau, 
\nonumber \\
&=&B(1,\eir)\hyg\left(1,\eir,1+\eir,z\right),\nonumber \\
&=&\int_0^1d\tau\biggl[\frac{\delta(\tau)}{\eir}(1-z\tau)^{-1}
+\frac{(1-z\tau)^{-1}-1}{\tau}
+\eir\frac{\ln\tau}{\tau}\left(\left(1-z\tau\right)^{-1}
-1\right)\nonumber \\
&+&{\cal O}\left(\eir^2\right)\biggr], \nonumber \\
&=&\frac{1}{\eir}-\ln(1-z)-\Li(z)\eir+{\cal O}\left(\eir^2\right). \label{hyg6} 
\end{eqnarray}
Then the first three terms of Eq.(\ref{hyg7}) are
\begin{eqnarray}
{\cal F}^{(-1)}_{0,0,0}(z)&=&1, \\
{\cal F}^{(0)}_{0,0,0}~(z)&=&-\ln(1-z), \\
{\cal F}^{(1)}_{0,0,0}~(z)&=&-\Li(z).
\end{eqnarray}
When $l=m=0,n\neq0$, 
the integral becomes
\begin{eqnarray}
{\cal I}_{0,0,n}&=&\int_0^1 \tau^{n-1+\eir}(1-z \tau)^{-1} d\tau.
\end{eqnarray}
Then, the first three terms of the Eq.(\ref{hyg7}) are
\begin{eqnarray}
{\cal F}^{(0)}_{0,0,n}(z)&=&\frac{\hyg(1,n,1+n;z)}{n}, \\
{\cal F}^{(1)}_{0,0,n}(z)&=&-{\tilde F}^{(1)}_{n,0}(z), \\
{\cal F}^{(2)}_{0,0,n}(z)&=&{\tilde F}^{(2)}_{n,0}(z).
\end{eqnarray}
When $l=0,m\neq0,n=0$,
the integral becomes
\begin{eqnarray}
{\cal I}_{0,m,0}&=&\int_0^1 \tau^{-1+\eir}(1-\tau)^m(1-z \tau)^{-1} d\tau \\
&=&\sum_{k=0}^{m} ~_mC_k (-1)^k \int_0^1 \tau^{k-1+\eir}(1-z \tau)^{-1} d\tau \\
&=&\sum_{k=0}^{m} ~_mC_k (-1)^k {\cal I}_{0,0,k}.
\end{eqnarray}
Then, the first three terms of the Eq.(\ref{hyg7}) are
\begin{eqnarray}
{\cal F}^{(-1)}_{0,m,0}(z)&=&{\cal F}^{(-1)}_{0,0,0}(z),\\
{\cal F}^{(0)}_{0,m,0}(z)~&=&\sum_{k=0}^{m} ~_mC_k (-1)^k {\cal F}^{(0)}_{0,0,k}(z),\\
{\cal F}^{(1)}_{0,m,0}(z)
~&=&\sum_{k=0}^{m} ~_mC_k (-1)^k {\cal F}^{(1)}_{0,0,k}(z).
\end{eqnarray}
When $l=0,m\neq0,n\neq0$,
the integral becomes
\begin{eqnarray}
{\cal I}_{0,m,n}&=&\int_0^1 \tau^{n-1+\eir}(1-\tau)^m(1-z \tau)^{-1} d\tau \\
&=&\sum_{k=0}^{m} ~_mC_k (-1)^k \int_0^1 \tau^{n+k-1+\eir}(1-z \tau)^{-1} d\tau \\
&=&\sum_{k=0}^{m} ~_mC_k (-1)^k {\cal I}_{0,0,n+k}.
\end{eqnarray}
Then, the first three terms of the Eq.(\ref{hyg7}) are
\begin{eqnarray}
{\cal F}^{(0)}_{0,m,n}(z)&=&B(1+m,n)\hyg(1,n,1+n+m;z),\\
{\cal F}^{(1)}_{0,m,n}(z)&=&\sum_{k=0}^{m} ~_mC_k (-1)^{k+1}
{\tilde F}^{(1)}_{n+k,0}(z), \\
{\cal F}^{(2)}_{0,m,n}(z)&=&\sum_{k=0}^{m} ~_mC_k (-1)^k 
{\tilde F}^{(2)}_{n+k,0}(z).
\end{eqnarray}
When $l\neq0,m=0,n=0$,
the integral becomes
\begin{eqnarray}
{\cal I}_{l,0,0}&=&\int_0^1 \tau^{l-1+\eir}(1-z \tau)^{-(l+1)} d\tau.
\end{eqnarray}
Then, the first three terms of the Eq.(\ref{hyg7}) are
\begin{eqnarray}
{\cal F}^{(0)}_{l,0,0}(z)&=&\frac{(1-z)^{-l}}{l}, \\
{\cal F}^{(1)}_{l,0,0}(z)&=&-\frac{\hyg(l,l,l+1;z)}{l^2},\\
{\cal F}^{(2)}_{l,0,0}(z)&=&{\tilde F}^{(2)}_{l,l}(z).
\end{eqnarray}
When $l\neq0,m=0,n\neq0$,
the integral becomes
\begin{eqnarray}
{\cal I}_{l,0,n}&=&\int_0^1 \tau^{l+n-1+\eir}(1-z \tau)^{-(l+1)} d\tau.
\end{eqnarray}
Then, the first three terms of the Eq.(\ref{hyg7}) are
\begin{eqnarray}
{\cal F}^{(0)}_{l,0,n}(z)&=&\frac{\hyg(1+l,l+n,1+l+n;z)}{l+n},\\
{\cal F}^{(1)}_{l,0,n}(z)&=&-{\tilde F}^{(1)}_{l+n,l}(z),\\
{\cal F}^{(2)}_{l,0,n}(z)&=&{\tilde F}^{(2)}_{l+n,l}(z).
\end{eqnarray}
When $l\neq0,m\neq0,n=0$,
the integral becomes
\begin{eqnarray}
{\cal I}_{l,m,0}&=&\int_0^1 \tau^{l-1+\eir}(1-\tau)^m(1-z \tau)^{-(l+1)} d\tau.
\end{eqnarray}
Then, the first three terms of the Eq.(\ref{hyg7}) are
\begin{eqnarray}
{\cal F}^{(0)}_{l,m,0}(z)&=&
B(l,1+m)\hyg(1+l,l,1+l+m;z),\\
{\cal F}^{(1)}_{l,m,0}(z)&=&
\sum_{k=0}^m~_mC_k(-1)^{k+1}{\tilde F}^{(1)}_{l+k,l}(z),\\
{\cal F}^{(2)}_{l,m,0}(z)&=&
\sum_{k=0}^m~_mC_k(-1)^{k}{\tilde F}^{(2)}_{l+k,l}(z).
\end{eqnarray}
When $l\neq0,m\neq0,n\neq0$,
the integral becomes
\begin{eqnarray}
{\cal I}_{l,m,n}&=&\int_0^1 
\tau^{l+n-1+\eir}(1-\tau)^m(1-z \tau)^{-(l+1)} d\tau.
\end{eqnarray}
Then, the first three terms of the Eq.(\ref{hyg7}) are
\begin{eqnarray}
{\cal F}^{(0)}_{l,m,n}&=&
B(l+n,1+m)\hyg(1+l,l+n,1+l+m+n;z),\\
{\cal F}^{(1)}_{l,m,n}&=&
\sum_{k=0}^m~_mC_k(-1)^{k+1}{\tilde F}^{(1)}_{l+n+k,l}(z),\\
{\cal F}^{(2)}_{l,m,n}&=&
\sum_{k=0}^m~_mC_k(-1)^{k}{\tilde F}^{(2)}_{l+n+k,l}(z).
\end{eqnarray}

\section{$1/\eir$ expansion of tensor-integrations 
with the $\overline{MS}$ scheme}
\setcounter{equation}{0}
We give $1/\eir$ expansions of tensor-integrals under the
$\overline{MS}$ scheme in this Appendix.  The 
$\overline{MS}$ scheme is realized with the following replacement of
the renormalization energy-scale:
\begin{eqnarray*}
\mu_R^2 \rightarrow \mu_R^2\frac{e^{\gamma_E}}{4\pi},
\end{eqnarray*}
where $\gamma_E$ is Euler's constant.
\subsection{Vertex with a 1 off-shell external line}
The general result of the vertex tensor-integral is shown in 
Eq.(\ref{v00p}). The IR structure of this function can be obtained
by a Laurant expansion of the beta function for the IR singular case.
\subsubsection{$n_x=n_y=0$}
When $n_x=n_y=0$, the tensor integral becomes
\begin{eqnarray}
J_{3}(0,0,p_3^2;0,0)&=&
\frac{\Gamma(-\eir)}{(4 \pi)^2}
\left(\frac{-\tps}{4 \pi \mu_R^2}\right)^{\eir}
\frac{1}{-p_3^2}
\frac{B(\eir,\eir)}{2}.
\end{eqnarray}
After the $\eir$ expansion with the $\overline{MS}$ scheme,
\begin{eqnarray}
J_{3}(0,0,p_3^2;0,0)&\rightarrow&\frac{1}{(4\pi)^2~p_3^2}
\left[\frac{\Ct_{31}}{\eir^2}+\frac{\Co_{31}}{\eir}+\Cz_{31}+O(\eir)\right],
\end{eqnarray}
where
\begin{eqnarray}
\Ct_{31}&=&1, \\
\Co_{31}&=&\ln\left(\frac{-p_3^2}{\mu_R^2}\right), \\
\Cz_{31}~&=&-\frac{\pi^2}{12}+
\frac{1}{2}\ln^2\left(\frac{-p_3^2}{\mu_R^2}\right).
\end{eqnarray}
\subsubsection{$n_x+n_y\neq0$}
The general result of the vertex tensor-integral, Eq.(\ref{v00p}), is
symmetric under $n_x$ and $n_y$ exchanges.
When one of $n_x$ or $n_y$ is non-zero, the tensor integral becomes
\begin{eqnarray}
J_{3}(0,0,p_3^2;n,0)&=&J_{3}(0,0,p_3^2;0,n)\nonumber \\
&=&
\frac{\eir \Gamma(-\eir)}{(4 \pi)^2}
\left(\frac{-\tps}{4 \pi \mu_R^2}\right)^{\eir}
\frac{1}{-p_3^2}
\frac{B(n+\eir,\eir)}{n+2\eir}.
\end{eqnarray}
After an $\eir$ expansion with the $\overline{MS}$ scheme,
\begin{eqnarray}
J_{3}(0,0,p_3^2;n,0)&=&J_{3}(0,0,p_3^2;0,n)\nonumber \\
&\rightarrow&\frac{1}{(4\pi)^2~p_3^2}
\left[\frac{\Co_{32}}{\eir}+\Cz_{32}+O(\eir)\right],
\end{eqnarray}
where
\begin{eqnarray}
\Co_{32}&=&\frac{1}{n}, \\
\Cz_{32}~&=&-\frac{2}{n^2}+\frac{1}{n}\left(\ln\left(\frac{-p_3^2}{\mu_R^2}\right)
-{\cal H}_{n-1}\right),
\end{eqnarray}
where ${\cal H}_m$ is the harmonic number, defined as
\begin{eqnarray*}
{\cal H}_{m}&=\sum_{j=1}^m \frac{1}{j},
\end{eqnarray*}
\subsubsection{$n_x\neq0, n_y\neq0$}
In this case, because there is no IR-divergence in Eq.(\ref{v00p}), 
$\eir$ can be set to zero,
\begin{eqnarray}
J_{3}(0,0,p_3^2;n_x,n_y) 
&\rightarrow&
\frac{1}{(4 \pi)^2 p_3^2 }\frac{(n_x-1)! (n_y-1)!}{(n_x+n_y)!}.
\end{eqnarray}
\subsection{Vertex with two off-shell external lines}
\subsubsection{$n_x=n_y=0$}
The function ${\cal G}_0(z)$, defined by Eq.(\ref{gfunc}),
can be expanded with respect to $\eir$ around zero as
\begin{eqnarray}
{\cal G}_0(z)&=&
{\cal G}^{(1)}_0(z)\eir+
{\cal G}^{(2)}_0(z)\eir^2
+{\cal O}\left(\eir^3\right),
\end{eqnarray}
where
\begin{eqnarray}
{\cal G}^{(1)}_0(z)&=& \frac{\ln(1-z)}{z}, \\
{\cal G}^{(2)}_0(z)&=& \frac{\ln^2(1-z)}{2z}.
\end{eqnarray}
Then, the vertex integral $J_3(0,p_2^2,p_3^2;0,0)$ can be express
after an $\eir$ expansion with the $\overline{MS}$ scheme as
\begin{eqnarray}
J_{3}(0,p_2^2,p_3^2;0,0)&\rightarrow&\frac{1}{(4\pi)^2~p_3^2}
\left[\frac{\Co_{33}}{\eir}+\Cz_{33}+O(\eir)\right],
\end{eqnarray}
where
\begin{eqnarray}
\Co_{33}&=&
\Ct_{31}{\cal G}^{(1)}_0\left(\frac{p_3^2-p_2^2}{\tps}\right), \\
\Cz_{33}~&=&
\Co_{31}{\cal G}^{(1)}_0\left(\frac{p_3^2-p_2^2}{\tps}\right)+
\Ct_{31}{\cal G}^{(2)}_0\left(\frac{p_3^2-p_2^2}{\tps}\right).
\end{eqnarray}
\subsubsection{$n_x=0, n_y\neq0$}
Because of 
vertex integral $J_3(0,p_2^2,p_3^2;0,n)$ is IR finite, we can set
$\eir$ to be zero as
\begin{eqnarray}
J_{3}(0,p_2^2,p_3^2;0,n)&=&\frac{1}{(4\pi)^2~p_3^2} \Cz_{34},
\end{eqnarray}
where
\begin{eqnarray}
\Cz_{34}~&=&
\Co_{32}{\cal G}_0^{(1)}\left(\frac{p_3^2-p_2^2}{\tps}\right).
\end{eqnarray}
\subsubsection{$n_x\neq0, n_y=0$}
For non-zero values of n, ${\cal G}_n(z)$ 
can be expanded with respect to $\eir$ around zero as
\begin{eqnarray}
{\cal G}_n(z)&=&
{\cal G}^{(0)}_n(z)+
{\cal G}^{(1)}_n(z)\eir
+{\cal O}\left(\eir^2\right),
\end{eqnarray}
where
\begin{eqnarray}
{\cal G}^{(0)}_n(z)&=& \frac{n}{n+1}\hyg(1,1,2+n;z),\\
{\cal G}^{(1)}_n(z)&=& \frac{1}{n+1}\hyg(1,1,2+n;z)
-n\left(\frac{z-1}{z}\right)^n\frac{\ln^2(1-z)}{2z} \nonumber \\
&+&\frac{n}{z^n}\sum_{k=1}^n~_nC_k (z-1)^{n-k}
\frac{(1-z)^k\left(1-k \ln(1-z)\right)-1}{k^2z}.
\end{eqnarray}
Then, the vertex integral $J_3(0,p_2^2,p_3^2;n,0)$ can be express
after an $\eir$ expansion with the $\overline{MS}$ scheme as
\begin{eqnarray}
J_{3}(0,p_2^2,p_3^2;n,0)&\rightarrow&\frac{1}{(4\pi)^2~p_3^2}
\left[\frac{\Co_{35}}{\eir}+\Cz_{35}+O(\eir)\right],
\end{eqnarray}
where
\begin{eqnarray}
\Co_{35}&=&
\Co_{32}{\cal G}^{(0)}_0\left(\frac{p_3^2-p_2^2}{\tps}\right), \\
\Cz_{35}~&=&
\Co_{32}{\cal G}^{(1)}_0\left(\frac{p_3^2-p_2^2}{\tps}\right)+
\Cz_{31}{\cal G}^{(0)}_0\left(\frac{p_3^2-p_2^2}{\tps}\right).
\end{eqnarray}
\subsubsection{$n_x\neq0, n_y\neq0$}
In this case, there is no IR-divergence and $\eir$ can be set to zero,
\begin{eqnarray}
J_{3}(0,p_2^2,p_3^2;n_x,n_y) &=&
\frac{1}{(4 \pi)^2 p_3^2 }\frac{(n_x-1)! (n_y-1)!}{(n_x+n_y)!}\nonumber \\
&\times&
\hyg\left(1,1,2+n_x;\frac{p_3^2-p_2^2}{\tps}\right)
\frac{n_x}{n_x+1}.
\end{eqnarray}
\subsection{Box integral with all on-shell external legs}
The general result of the box tensor-integral 
with all on-shell external legs
is shown in 
Eq.(\ref{b0-nnn}). The IR structure of this function can be obtained
by a Laurant expansion of the beta function for the IR singular case.
\subsubsection{$n_x=n_y=n_z=0$}
When the numerator of the integrand is unity, $n_x=n_y=n_z=0$,
the box tensor-integral result, Eq.(\ref{b0-nnn}), is reduced to
\begin{eqnarray}
&~&J_4(s,t,0,0,0,0;0,0,0)=\frac{1}{(4\pi)^2s~t}\frac{B(\eir,\eir)\Gamma(1-\eir)}{\eir}
\nonumber \\
&\times&\left[\left(\frac{-\s}{4 \pi \mu_R^2}\right)^{\eir}
\hyg\left(1,\eir,1+\eir,-\frac{\u}{\bar t}\right)
+\left(\frac{-\t}{4 \pi \mu_R^2}\right)^{\eir}
\hyg\left(1,\eir,1+\eir,-\frac{\u}{\s}\right)\right].\nonumber \\
\end{eqnarray}
After using a Laurent expansion of the hypergeometric function 
of Eq.(\ref{hyg6}) the loop integral with the $\overline{MS}$ scheme is 
\begin{eqnarray}
J_4(s,t,0,0,0,0;0,0,0)
&\rightarrow&\frac{1}{(4 \pi)^2s~t}
\left[
\frac{\Ct_{41}}{\eir^2}+\frac{\Co_{41}}{\eir}+\Cz_{41}
+O(\eir)\right], \\
\Ct_{41}&=&4,\\
\Co_{41}&=&2\left[\ln\left(\frac{-\s}{\mu_R^2}\right)
+\ln\left(\frac{-\t}{\mu_R^2}\right)\right],\\
\Cz_{41}~&=&-\frac{\pi^2}{3}
-2{\rm Li}_2\left(-\frac{\u}{\s}\right)
-2{\rm Li}_2\left(-\frac{\u}{\bar t}\right)
+\ln^2\left(\frac{-\s}{\mu_R^2}\right)
+\ln^2\left(\frac{-\t}{\mu_R^2}\right) \nonumber \\
&~&-2\ln\left(\frac{-\s}{\mu_R^2}\right) \ln\left(1+\frac{\u}{\bar t}\right)
-2\ln\left(\frac{-\t}{\mu_R^2}\right) \ln\left(1+\frac{\u}{\s}\right).
\end{eqnarray}
\subsubsection{$n_x=0,n_y=0,n_z\neq0$}
When $n_x=n_y=0,n_z\neq0$,
the result is reduced to
\begin{eqnarray}
&~&J_4(s,t;0,0,0,0;0,0,n_z)=\frac{B(\eir,n_z+\eir)\Gamma(1-\eir)}{(4\pi)^2s~t}
\nonumber \\
&\times&\biggl[\left(\frac{-\s}{4 \pi \mu_R^2}\right)^{\eir}
B(1,n_z+\eir)
\hyg\left(1,n_z+\eir,1+n_z+\eir,-\frac{\u}{\bar t}\right)\nonumber \\
&+&~\left(\frac{-\t}{4 \pi \mu_R^2}\right)^{\eir}
B(\eir,1+n_z)
\hyg\left(1,\eir,1+n_z+\eir,-\frac{\u}{\s}\right)\biggr]. \label{hyg2}
\end{eqnarray}
Laurent expansions of beta- and gamma-function
with the $\overline{MS}$ scheme
are obtained as follows:
\begin{eqnarray}
B(n_1+\eir,n_2+\eir)\Gamma(1-\eir)\left(\frac{-q^2}{4\pi\mu_R^2}\right)^{\eir}
&\rightarrow&
\frac{{\cal A}^{(-1)}_{n_1,n_2}(q^2)}{\eir}+{\cal A}^{(0)}_{n_1,n_2}(q^2)
+{\cal A}^{(1)}_{n_1,n_2}(q^2)\eir, \nonumber \\
\end{eqnarray}
where
\def\LQ{\ln\left(\frac{-q^2}{\mu^2_R}\right)}
\begin{eqnarray}
{\cal A}^{(-1)}_{0,n}(q^2)&=&1,\\
{\cal A}^{(0)}_{0,n}(q^2)~&=&\LQ-{\cal H}_{n-1}, \\
{\cal A}^{(1)}_{0,n}(q^2)~&=&\frac{1}{6}\biggl[
3{\cal H}_{n-1}^2+\pi^2+3\LQ\left(\LQ-2{\cal H}_{n}\right)
-9\psi^{(1)}(n)
\biggr], 
\end{eqnarray}
and
\begin{eqnarray}
{\cal A}^{(-1)}_{n_1,n_2}(q^2)&=&0,\\
{\cal A}^{(0)}_{n_1,n_2}(q^2)~&=&B(n_1,n_2), \\
{\cal A}^{(1)}_{n_1,n_2}(q^2)~&=&B(n_1,n_2)
\left({\cal H}_{n_1-1}+{\cal H}_{n_2}-2{\cal H}_{n_1+n_2-1}
+\ln\left(\frac{-q^2}{\mu^2_R}\right)\right),
\end{eqnarray}
where $n,~n_1,~n_2$ are positive integers.
Here, $\psi^{(1)}(z)$ is the first derivative of the digamma function, given by
\begin{eqnarray*}
\psi^{(1)}(n)&=&\frac{\pi^2}{6}-\sum_{k=1}^{n-1}\frac{1}{k^2}.
\end{eqnarray*}

Finally, the tensor integral can be obtained to be
\def\fracs{\left(-\frac{\u}{\s}\right)}
\def\fract{\left(-\frac{\u}{\tb}\right)}
\def\A{{\cal A}_{0,n_z}}
\def\B{{\cal F}}
\begin{eqnarray}
J_4(s,t;0,0,0,0;0,0,n_z)&\rightarrow&\frac{1}{(4\pi)^2s~t}
\left[
\frac{\Ct_{42}}{\eir^2}+\frac{\Co_{42}}{\eir}+\Cz_{42}
+O(\eir)\right], 
\end{eqnarray}
where
\begin{eqnarray}
\Ct_{42}&=&\A^{(-1)}(\t)\B^{(-1)}_{0,n_z,0}\fracs, \\
\Co_{42}&=&
\A^{(-1)}(\s)\B^{(0)}_{0,0,n_z}\fract\nonumber \\
&+&\A^{(-1)}(\t)\B^{(0)}_{0,n_z,0}\fracs
+\A^{(0)}(\t)\B^{(-1)}_{0,n_z,0}\fracs, \\
\Cz_{42}~&=&
\A^{(-1)}(\s)\B^{(1)}_{0,0,n_z}\fract
+\A^{(0)}(\s)\B^{(0)}_{0,0,n_z}\fract\nonumber \\
&+&\A^{(-1)}(\t)\B^{(1)}_{0,n_z,0}\fracs
+\A^{(0)}(\t)\B^{(0)}_{0,n_z,0}\fracs\nonumber \\
&+&\A^{(1)}(\t)\B^{(-1)}_{0,n_z,0}\fracs.
\end{eqnarray}
\subsubsection{$n_x=0,n_y\neq0,n_z=0$}
When $n_x=0$ and both $n_y$ and $n_z$ are non zero, $J_4$ becomes
\begin{eqnarray}
&~&J_4(s,t;0,0,0,0;0,n_y,n_z)=\frac{1}{(4\pi)^2s~t}B(\eir,n_y+n_z+\eir)\Gamma(1-\eir)
\nonumber \\
&\times&
\biggl[\left(\frac{-\s}{4 \pi \mu_R^2}\right)^{\eir}
B(1+n_y,n_z+\eir)
\hyg\left(1,n_z+\eir,1+n_y+n_z+\eir,-\frac{\u}{\bar t}\right)\nonumber \\
&+&~\left(\frac{-\t}{4 \pi \mu_R^2}\right)^{\eir}
B(1+n_z,n_y+\eir)
\hyg\left(1,n_y+\eir,1+n_y+n_z+\eir,-\frac{\u}{\s}\right)
\biggr].
\nonumber \label{0nn}\\
\end{eqnarray}
One can see that $J_4(s,t;0,0,0,0;0,n_y,n_z)$ is symmetric under a
simultaneous exchange of $t \leftrightarrow s, n_y \leftrightarrow  n_z$.
Then, for the case of $n_x=n_z=0,n_y\neq0$, the required integration can be
obtained as
\begin{eqnarray}
J_4(s,t;0,0,0;0,n_y,0)&=&\{J_4(s,t;0,0,0;0,0,n_z),n_z \rightarrow n_y,
s\leftrightarrow t\}.
\end{eqnarray}
\subsubsection{$n_x=0,n_y\neq0,n_z\neq0$}
Let's start from eq.(\ref{0nn}). There is no IR-singular pole in
the square-bracket and the $\frac{1}{\eir}$ pole in the Beta function
in front of the square brackets. It is thus needed to expand
the terms in the square-bracket up to $O\left(\eir\right)$.

Then, finally we obtain
\begin{eqnarray}
&~&J_4(s,t;0,0,0,0;0,n_y,n_z)=
\frac{1}{(4\pi)^2s~t}B(\eir,n_y+n_z+\eir)\Gamma(1-\eir)
\nonumber \\
&\times&
\biggl[
\left(\frac{-\s}{4 \pi \mu_R^2}\right)^{\eir}\left(
{\cal F}^{(0)}_{0,n_y,n_z}\left(-\frac{\u}{\tb}\right)
+\eir
{\cal F}^{(1)}_{0,n_y,n_z}\left(-\frac{\u}{\tb}\right)
\right)
\nonumber \\
&+&~\left(\frac{-\t}{4 \pi \mu_R^2}\right)^{\eir}\left(
{\cal F}^{(0)}_{0,n_z,n_y}\left(-\frac{\u}{\s}\right)
+\eir
{\cal F}^{(1)}_{0,n_z,n_y}\left(-\frac{\u}{\s}\right)
\right)
\biggr].
\end{eqnarray}

After an $\eir$ expansion with the $\overline{MS}$ scheme,
it is obtained as
\begin{eqnarray}
J_4(s,t;0,0,0,0;0,n_y,n_z)&\rightarrow&\frac{1}{(4\pi)^2s~t}
\left[\frac{\Co_{43}}{\eir}+\Cz_{43}+O(\eir)\right], 
\end{eqnarray}
where
\begin{eqnarray}
\Co_{43}&=&
{\cal F}^{(0)}_{0,n_y,n_z}\left(-\frac{\u}{\tb}\right)+
{\cal F}^{(0)}_{0,n_z,n_y}\left(-\frac{\u}{\s}\right),\\
\Cz_{43}~&=&
{\cal F}^{(1)}_{0,n_y,n_z}\left(-\frac{\u}{\tb}\right)+
{\cal F}^{(1)}_{0,n_z,n_y}\left(-\frac{\u}{\s}\right)\nonumber \\
&+&{\cal F}^{(0)}_{0,n_y,n_z}\left(-\frac{\u}{\tb}\right)
\left(\ln\left(-\frac{\s}{\mu_R^2}\right)-{\cal H}_{n_y+n_z-1}\right)\nonumber \\
&+&{\cal F}^{(0)}_{0,n_z,n_y}\left(-\frac{\u}{\s}\right)
\left(\ln\left(-\frac{\t}{\mu_R^2}\right)-{\cal H}_{n_y+n_z-1}\right).
\end{eqnarray}
\subsubsection{$n_x\neq0,n_y=0,n_z=0$}
The basic integrand, Eq.(\ref{box4onshell}), is symmetric under the
exchange $x(n_x) \leftrightarrow z(n_z)$. Then, the result can be 
easily obtained as
\begin{eqnarray}
J_4(s,t;0,0,0,0;n_x,0,0)&=&\{J_4(s,t;0,0,0,0;0,0,n_z),n_z \rightarrow n_x\}.
\end{eqnarray}
\subsubsection{$n_x\neq0,n_y\neq0,n_z=0$}
When $n_z=0$, the result of the integration, Eq.(\ref{b0-nnn}), can be
\begin{eqnarray}
&~&J_4(s,t;0,0,0,0;n_x,n_y,0)
=\frac{1}{(4\pi)^2s~t}B(n_x+\eir,n_y+\eir)n_x! \Gamma(\eir) \Gamma(1-\eir)
\nonumber \\
&\times&\biggl[
\left(\frac{-\t}{4 \pi \mu_R^2}\right)^{\eir}
\left(\frac{-t}{s}\right)^{n_x} 
\frac{B(1+n_x+n_y+\eir)}{\Gamma(n_x+\eir)}
\nonumber \\
&\times&\hyg\left(1+n_x,n_x+n_y+\eir,1+n_x+n_y+\eir,-\frac{\u}{\s}\right)
\nonumber \\
&+&\left(\frac{-\s}{4 \pi \mu_R^2}\right)^{\eir}
\sum_{l=0}^{n_x} \left(\frac{-s}{t}\right)^l 
\frac{(-1)^l}{\Gamma(l+\eir)(n_x-l)!}
B(1+n_y,l+\eir)
\nonumber \\
&\times&\hyg\left(1+l,l+\eir,1+l+n_y+\eir,-\frac{\u}{\bar t}\right)
\biggr]. 
\end{eqnarray}

After an $\eir$ expansion with the $\overline{MS}$ scheme,
it is obtained as
\begin{eqnarray}
J_4(s,t;0,0,0,0;n_x,n_y,0)&\rightarrow&\frac{1}{(4\pi)^2s~t}
\left[\frac{\Co_{44}}{\eir}+\Cz_{44}+O(\eir)\right], 
\end{eqnarray}
\def\fmxy{{\cal F}^{(-1)}_{n_x,0,n_y}}
\def\fzxy{{\cal F}^{(0)}_{n_x,0,n_y}}
\def\fpxy{{\cal F}^{(1)}_{n_x,0,n_y}}
\def\fmy{{\cal F}^{(-1)}_{0,n_y,0}}
\def\fzy{{\cal F}^{(0)}_{0,n_y,0}}
\def\fpy{{\cal F}^{(1)}_{0,n_y,0}}
\def\fmly{{\cal F}^{(-1)}_{l,n_y,0}}
\def\fzly{{\cal F}^{(0)}_{l,n_y,0}}
\def\fply{{\cal F}^{(1)}_{l,n_y,0}}
\def\aazt{{\cal A}^{(0)}_{n_x,n_y}(\t)}
\def\aapt{{\cal A}^{(1)}_{n_x,n_y}(\t)}
\def\aazs{{\cal A}^{(0)}_{n_x,n_y}(\s)}
\def\aaps{{\cal A}^{(1)}_{n_x,n_y}(\s)}
\def\ess{\left(-\frac{\u}{\s}\right)}
\def\est{\left(-\frac{\u}{\tb}\right)}
where
\begin{eqnarray}
\Co_{44}&=&n_x~\left(-\frac{t}{s}\right)^{n_x}
\aazt \fzxy \ess+ \aazs \fmy \est \nonumber \\
&+&\sum_{l=1}^{n_x} l \left(\frac{s}{t}\right)^l~_{n_x}C_{l}~\aazs \fzly \est, 
\end{eqnarray}
\begin{eqnarray}
\Cz_{44}~&=&n_x~\left(-\frac{t}{s}\right)^{n_x}
\biggl[\aapt \fzxy \ess + \aazt \fpxy \ess \nonumber \\
&~&~~~~-\aazt \fzxy \ess {\cal H}_{n_x-1} \biggr] \nonumber \\
&+& \aaps \fmy \est + \aazs \fzy \est \nonumber \\
&+&\sum_{l=1}^{n_x} l \left(\frac{s}{t}\right)^l~_{n_x}C_{l}
\biggl[\aaps \fzly \est 
\nonumber \\
&+& \aazs\left(\fply \est - \fzly \est {\cal H}_{l-1}
\right)\biggr].
\end{eqnarray}
\subsubsection{$n_x\neq0,n_y=$any,$n_z\neq0$}
In the case of $n_x\neq0,n_z\neq0$, the integration has no
IR-divergence with any value of $n_y\geq0$.
The result is given by 
\begin{eqnarray}
J_4(s,t;0,0,0,0;n_x,n_y,n_z)&\rightarrow&\frac{1}{(4\pi)^2s~t}
\left[\Cz_{45}+O(\eir)\right], 
\end{eqnarray}
\def\fmxz{{\cal F}^{(-1)}_{n_x,n_z,n_y}}
\def\fzxz{{\cal F}^{(0)}_{n_x,n_z,n_y}}
\def\fpxz{{\cal F}^{(1)}_{n_x,n_z,n_y}}
\def\fmz{{\cal F}^{(-1)}_{0,n_y,n_z}}
\def\fzz{{\cal F}^{(0)}_{0,n_y,n_z}}
\def\fpz{{\cal F}^{(1)}_{0,n_y,n_z}}
\def\fmlz{{\cal F}^{(-1)}_{l,n_y,n_z}}
\def\fzlz{{\cal F}^{(0)}_{l,n_y,n_z}}
\def\fplz{{\cal F}^{(1)}_{l,n_y,n_z}}
\def\aazt{{\cal A}^{(0)}_{n_x,n_y+n_z}(\t)}
\def\aapt{{\cal A}^{(1)}_{n_x,n_y+n_z}(\t)}
\def\aazs{{\cal A}^{(0)}_{n_x,n_y+n_z}(\s)}
\def\aaps{{\cal A}^{(1)}_{n_x,n_y+n_z}(\s)}
\def\ess{\left(-\frac{\u}{\s}\right)}
\def\est{\left(-\frac{\u}{\tb}\right)}
where
\begin{eqnarray}
\Cz_{45}~&=&n_x~\left(-\frac{t}{s}\right)^{n_x}
\biggl[\aapt \fzxz \ess \nonumber \\
&+& \aazt\left(\fpxz \ess - \fzxz \ess {\cal H}_{n_x-1}\right)
\biggr] \nonumber \\
&+& \aazs \fzz \est  \nonumber \\
&+&\sum_{l=1}^{n_x} l \left(\frac{s}{t}\right)^l~_{n_x}C_{l}
\biggl[\aaps \fzlz \est 
\nonumber \\
&+& \aazs\left(\fplz \est - \fzlz \est {\cal H}_{l-1}
\right)\biggr].
\label{res1}
\end{eqnarray}
\subsection{Box integral with 1 off-shell and 3 on-shell external legs}
The general result of the box tensor-integral 
with 1 off-shell and 3 on-shell external legs
is shown in 
Eq.(\ref{b1-nnn}) with Eqs.(\ref{b1-nnn-I1})
and (\ref{b1-nnn-I2}). The IR structure of this function can be obtained
by the Laurant expansion of the beta function for the IR singular case.
\subsubsection{$n_x=n_y=n_z=0$}
When the numerator of the integrand is unity, $n_x=n_y=n_z=0$,
the second integral $cal{I}^{\rm (2)}$ is reduced to
\begin{eqnarray}
\cal{I}^{\rm (2)}&=&\int_0^1 dw~
\left(1+\frac{\u}{\s}w\right)^{-1}\left(1+\frac{\t+\u}{\s}w\right)^{\eir-1} \\
&=&\frac{-s}{t}\frac{1}{\eir}\left[
\hyg\left(1,\eir,1+\eir,-\frac{\u}{\bar t}\right)
-\left(\frac{{\tilde{p}_1^2}}{\s}\right)^{\eir}
\hyg\left(1,\eir,1+\eir,-\frac{\u {\tilde{p}_1^2}}{{\bar t}\s}\right)
\right], \nonumber \\
\end{eqnarray}
where ${\tilde{p}_1^2}=p_1^2+i0$. 
Then, the final result can be written as
\begin{eqnarray}
&~&J_4(s,t;p_1^2,0,0,0;0,0,0)=\frac{1}{(4\pi)^2s~t}
\frac{B(\eir,\eir)\Gamma(1-\eir)}{\eir}
\nonumber \\
&\times&\biggl[\left(\frac{-\s}{4 \pi \mu_R^2}\right)^{\eir}
\hyg\left(1,\eir,1+\eir,-\frac{\u}{\bar t}\right)
+\left(\frac{-\t}{4 \pi \mu_R^2}\right)^{\eir}
\hyg\left(1,\eir,1+\eir,-\frac{\u}{\s}\right)\nonumber \\
&-&\left(\frac{-{\tilde{p}_1^2}}{4 \pi \mu_R^2}\right)^{\eir}
\hyg\left(1,\eir,1+\eir,-\frac{\u {\tilde{p}_1^2}}{{\bar t}\s}\right)
\biggr], \\
&=&J_4(s,t;0,0,0,0;0,0,0)\nonumber \\
&-&\frac{1}{(4\pi)^2s~t}
\frac{B(\eir,\eir)\Gamma(1-\eir)}{\eir}
\left(\frac{-{\tilde{p}_1^2}}{4 \pi \mu_R^2}\right)^{\eir}
\hyg\left(1,\eir,1+\eir,-\frac{\u {\tilde{p}_1^2}}{{\bar t}\s}\right)
\biggr]. \nonumber \\
\end{eqnarray}
After an $\eir$ expansion with the $\overline{MS}$ scheme,
it is obtained as
\begin{eqnarray}
J_4(s,t,p_1^2,0,0,0;0,0,0)
&\rightarrow&\frac{1}{(4 \pi)^2s~t}\left[
\frac{\Ct_{46}}{\eir^2}+\frac{\Co_{46}}{\eir}+\Cz_{46}
+O(\eir)\right], 
\end{eqnarray}
where
\begin{eqnarray}
\Ct_{46}&=&\Ct_{41}-2,\\
\Co_{46}&=&\Co_{41}-2\left[
\ln\left(\frac{-\tilde{p}_1^2}{\mu_R^2}\right)
-\ln\left(1+\frac{\u \tpo}{\tb \s}\right)
\right],\\
\Cz_{46}~&=&\Cz_{41}+\biggl[\frac{\pi^2}{6}
+2{\rm Li}_2\left(-\frac{\u \tilde{p}_1^2}{{\bar t}\s}\right)
-\ln^2\left(\frac{- \tilde{p}_1^2}{\mu_R^2}\right)
+2\ln\left(\frac{-\tilde{p}_1^2}
{\mu_R^2}\right) \ln\left(1+\frac{\u\tilde{p}_1^2}
{{\bar t}\s}\right)\biggl].
\end{eqnarray}
\subsubsection{$n_x=0,n_y+n_z\geq1$}
When $n_x=0$, and $n_y,n_z$ have any value with $n_y+n_z\geq1$,
the final result can be written as
\begin{eqnarray}
J_4(s,t,p_1^2,0,0,0;0,n_y,n_z)
&=&\frac{1}{(4 \pi)^2s~t}
B(\eir,n_y+n_z+\eir)\Gamma(1-\eir)\nonumber \\
&\times&\left[ 
\left(\frac{-\t}{4 \pi \mu^2}\right)^{\eir}{\cal I}^{(1)}+
\left(\frac{-\s}{4 \pi \mu^2}\right)^{\eir}{\cal I}^{(2)}
\right], 
\end{eqnarray}
The integrals, $\cal{I}^{\rm{(1)}}$ and $\cal{I}^{\rm{(2)}}$ can be done
easily as
\begin{eqnarray}
\cal{I}^{\rm (1)}&=&B(n_y+\eir,1+n_z) 
\hyg\left(1,n_y+\eir,1+n_y+n_z+\eir,-\frac{\u}{\s}
\right) \\
{\cal I}^{\rm (2)}_0
&=&\sum_{k_1=0}^{n_z}
\sum_{k_2=0}^{n_y+k_1} 
~_{n_z}C_{k_1}~
~_{n_y+k_1}C_{k_2}~ 
(-1)^{k_1+k_2}~
\left(\frac{s}{s-p_1^2}\right)^{n_y+k_1}
\frac{1}{k_2+\eir}
\nonumber \\
&\times&\biggl[\hyg\left(1,k_2+\eir,1+k_2+\eir,-\frac{\u}{\bar t}\right)
-\left(\frac{\tpo}{\s}\right)^{\eir}
\hyg\left(1,k_2+\eir,1+k_2+\eir,-\frac{\u \tpo}{{\tb}\s}\right)
\biggr]. 
\nonumber \\
\end{eqnarray}
After an $\eir$ expansion with the $\overline{MS}$ scheme,
it is obtained as
\begin{eqnarray}
J_4(s,t,p_1^2,0,0,0;0,n_y,n_z)
&\rightarrow&\frac{1}{(4 \pi)^2s~t}\left[
\frac{\Ct_{47}}{\eir^2}+\frac{\Co_{47}}{\eir}+\Cz_{47}
+O(\eir)\right], 
\end{eqnarray}
where
\def\fracp{\left(\frac{p_1^2}{s}\right)}
\def\fmznz{{\cal F}^{(-1)}_{0,n_z,n_y}}
\def\fzznz{{\cal F}^{(0)}_{0,n_z,n_y}}
\def\foznz{{\cal F}^{(1)}_{0,n_z,n_y}}
\def\fmzzkt{{\cal F}^{(-1)}_{0,0,k_2}}
\def\fzzzkt{{\cal F}^{(0)}_{0,0,k_2}}
\def\fozzkt{{\cal F}^{(1)}_{0,0,k_2}}
\def\ess{\left(-\frac{\u}{\s}\right)}
\def\esto{\left(-\frac{\u}{\tb}\right)}
\def\estt{\left(-\frac{\u\tpo}{\tb\s}\right)}
\def\amot{{\cal A}^{(-1)}_{0,n_y+n_z}(\t)}
\def\azt{{\cal A}^{(0)}_{0,n_y+n_z}(\t)}
\def\aot{{\cal A}^{(1)}_{0,n_y+n_z}(\t)}
\def\amos{{\cal A}^{(-1)}_{0,n_y+n_z}(\s)}
\def\azs{{\cal A}^{(0)}_{0,n_y+n_z}(\s)}
\def\aos{{\cal A}^{(1)}_{0,n_y+n_z}(\s)}
\def\lnaa{\ln\left(\frac{\tpo}{\s}\right)}
\begin{eqnarray}
\Ct_{47}&=&\amot \fmznz\ess, 
\\
\Co_{47}&=&  
\amot\fzznz\ess
+\azt\fmznz\ess
\nonumber \\
&+&\sum_{k_1=0}^{n_z}\sum_{k_2=0}^{n_y+k_1}
~_{n_z}C_{k_1}
~_{n_y+k_1}C_{k_2}(-1)^{k_1+k_2}
\left(\frac{s}{s-p_1^2}\right)^{n_y+k_1}
\amos \nonumber \\
&~&\times\left[\fzzzkt\esto-\fracp^{k_2}\left(
\fzzzkt\estt+\lnaa\fmzzkt\estt
\right)\right], 
\\
\Cz_{47}~&=& 
\amot\foznz\ess
+\azt\fzznz\ess
+\aot\fmznz\ess
\nonumber \\
&+&\sum_{k_1=0}^{n_z}\sum_{k_2=0}^{n_y+k_1}
~_{n_z}C_{k_1}
~_{n_y+k_1}C_{k_2}(-1)^{k_1+k_2}
\left(\frac{s}{s-p_1^2}\right)^{n_y+k_1}\nonumber \\
&~&\times\biggl[
\amos\left(\fozzkt\esto-\fracp^{k_2}
\left(\fozzkt\estt+\lnaa \fzzzkt\estt\right)\right)
\nonumber \\ 
&~&+\azs\left(\fzzzkt\esto-\fracp^{k_2}
\left(\fzzzkt\estt+\lnaa \fmzzkt\estt\right)\right)
\biggr].\nonumber \\
\end{eqnarray}
\subsubsection{$n_x\neq0,n_y=0,n_z=0$}
When $n_x\neq0$, and $n_y=n_z=0$,
the final result can be written as
\begin{eqnarray}
&~&J_4(s,t;p_1^2,0,0,0;n_x,0,0) =
\frac{1}{(4\pi)^2s~t}B(n_x+\eir,\eir)n_x! \Gamma(\eir) \Gamma(1-\eir)
\nonumber \\
&\times&\biggl[
\left(\frac{-\t}{4 \pi \mu_R^2}\right)^{\eir}
\left(\frac{-t}{s}\right)^{n_x} 
\frac{1}{\Gamma(n_x+\eir)}\Io 
+\left(\frac{-\s}{4 \pi \mu_R^2}\right)^{\eir} 
\sum_{l=0}^{n_x} 
\frac{(-1)^l}{\Gamma(l+\eir)(n_x-l)!}\It
\biggr].\nonumber \\
\end{eqnarray}
The integrals, $\cal{I}^{\rm{(1)}}$ and $\cal{I}^{\rm{(2)}}$ can be done
easily as
\begin{eqnarray}
\cal{I}^{\rm (1)}&=&B(n_x+\eir,1) 
\hyg\left(1+n_x,n_x+\eir,1+n_x+\eir,-\frac{\u}{\s}
\right), \\
{\cal I}^{\rm (2)}_0
&=&\frac{1}{\eir}
\biggl[\hyg\left(1+l,l+\eir,1+l+\eir,-\frac{\u}{\bar t}\right)
-\left(\frac{\tpo}{\s}\right)^{\eir}
\hyg\left(1+l,l+\eir,1+l+\eir,-\frac{\u \tpo}{{\tb}\s}\right)
\biggr]. 
\nonumber \\
\end{eqnarray}
After an $\eir$ expansion with the $\overline{MS}$ scheme,
it is obtained as
\begin{eqnarray}
J_4(s,t,p_1^2,0,0,0;n_x,0,0)
&\rightarrow&\frac{1}{(4 \pi)^2s~t}\left[
\frac{\Ct_{48}}{\eir^2}+\frac{\Co_{48}}{\eir}+\Cz_{48}
+O(\eir)\right], 
\end{eqnarray}
where
\def\fmzzz{{\cal F}^{(-1)}_{0,0,0}}
\def\fzzzz{{\cal F}^{(0)}_{0,0,0}}
\def\fpzzz{{\cal F}^{(1)}_{0,0,0}}
\def\fmlzz{{\cal F}^{(-1)}_{l,0,0}}
\def\fzlzz{{\cal F}^{(0)}_{l,0,0}}
\def\fplzz{{\cal F}^{(1)}_{l,0,0}}
\def\ftlzz{{\cal F}^{(2)}_{l,0,0}}
\def\fmxzz{{\cal F}^{(-1)}_{n_x,0,0}}
\def\fzxzz{{\cal F}^{(0)}_{n_x,0,0}}
\def\fpxzz{{\cal F}^{(1)}_{n_x,0,0}}
\def\ftxzz{{\cal F}^{(2)}_{n_x,0,0}}
\def\axmot{{\cal A}^{(-1)}_{0,n_x}(\t)}
\def\axzt{{\cal A}^{(0)}_{0,n_x}(\t)}
\def\axot{{\cal A}^{(1)}_{0,n_x}(\t)}
\def\axmos{{\cal A}^{(-1)}_{0,n_x}(\s)}
\def\axzs{{\cal A}^{(0)}_{0,n_x}(\s)}
\def\axos{{\cal A}^{(1)}_{0,n_x}(\s)}
\begin{eqnarray}
\Ct_{48}&=&n_x\left(-\frac{t}{s}\right)^{n_x}\axmot\fzxzz\fracs+\axmos
\nonumber \\
&~&\times\left[\fmzzz\fract-\fmzzz\estt+\sum_{l=1}^{n_x}l(-1)^l
~_{n_x}C_l\left(
\fzlzz\est-\fzlzz\estt\right)\right]
\\
\Co_{48}&=&  
n_x \left(-\frac{t}{s}\right)^{n_x}
\left[\axmot\fpxzz\ess+\axzt\fzxzz\ess
-{\cal H}_{n_x-1}\axmot\fzxzz\ess\right]\nonumber \\
&+&\axmos\left[\fzzzz\est-\fzzzz\estt-\lnaa\fmzzz\estt\right]
\nonumber \\
&+&\axzs\left[\fmzzz\est-\fmzzz\estt\right]
\nonumber \\
&+&\sum_{l=1}^{n_x}l(-1)^l~_{n_x}C_l \biggl[\axmos\left(
\fplzz\est-\fplzz\estt-\lnaa\fzlzz\estt\right)
\nonumber \\
&~&+\left(\axzs-\axmos{\cal H}_{l-1}\right)\left(\fzlzz\est-\fzlzz\estt\right)
\biggr]
\\
\Cz_{48}~&=& 
n_x\left(-\frac{t}{s}\right)^{n_x}\biggl[
\axmot\ftxzz\ess+\axzt\fpxzz\ess+\axot\fzxzz\ess
\nonumber \\
&~&-{\cal H}_{n_x-1}\left(\axmot\fpxzz\ess+\axzt\fzxzz\ess\right)
\nonumber \\
&~&+\frac{1}{2}\left({\cal H}_{n_x-1}+\frac{\pi^2}{6}-\psi^{(1)}_{n_x}
\right)\axmos\fzxzz\ess\biggr]
\nonumber \\
&+&\axmos\left[\fpzzz\est-\fpzzz\estt-\lnaa\fzzzz\estt\right]
\nonumber \\
&+&\axzs \left[\fzzzz\est-\fzzzz\estt-\lnaa\fmzzz\estt\right]
\nonumber \\
&+&\axos \left[\fmzzz\est-\fmzzz\estt\right]
\nonumber \\
&+&\sum_{l=1}^{n_x}l(-1)^l~_{n_x}C_l\biggl[\axmos\left(
\ftlzz\est-\ftlzz\estt-\lnaa\fplzz\estt\right)
\nonumber \\
&~&+
\left(\axzs-\axmos{\cal H}_{l-1}\right)
\left(\fplzz\est-\fplzz\estt-\lnaa\fzlzz\estt\right)
\nonumber \\
&~&+\left(\axos-\axzs{\cal H}_{l-1}+
\axmos\frac{1}{2}\left({\cal H}_{l-1}+\frac{\pi^2}{6}-\psi^{(1)}_l\right)
\right)
\nonumber \\
&~&~~\times \left(\fzlzz\est-\fzlzz\estt\right)\biggr]
\end{eqnarray}
\subsubsection{$n_x\neq0,n_y+n_z\geq1$}
At last we treat the most general case with $n_x\neq0$, and $n_y+n_z\geq1$.
The final result can be written as
\begin{eqnarray}
&~&J_4(s,t;p_1^2,0,0,0;n_x,n_y,n_z) =
\frac{1}{(4\pi)^2s~t}B(n_x+\eir,n_y+n_z+\eir)n_x! \Gamma(\eir) \Gamma(1-\eir)
\nonumber \\
&\times&\biggl[
\left(\frac{-\t}{4 \pi \mu_R^2}\right)^{\eir}
\left(\frac{-t}{s}\right)^{n_x} 
\frac{1}{\Gamma(n_x+\eir)}\Io 
+\left(\frac{-\s}{4 \pi \mu_R^2}\right)^{\eir} 
\sum_{l=0}^{n_x} 
\frac{(-1)^l}{\Gamma(l+\eir)(n_x-l)!}\It
\biggr], \nonumber \\
\end{eqnarray}
where
\begin{eqnarray}
\Io&=&B(1+n_z,n_x+n_y+\eir)
\hyg\left(1+n_x,n_x+n_y+\eir,1+n_x +n_y+n_z+\eir,-\frac{\u}{\s}\right), 
\nonumber \\ \\
\It&=&\sum_{k_1=0}^{n_z}
\sum_{k_2=0}^{n_y+k_1} 
~_{n_z}C_{k_1}~
~_{n_y+k_1}C_{k_2}~ 
(-1)^{k_1+k_2}~
\left(\frac{s}{s-p_1^2}\right)^{n_y+k_1}
\frac{1}{l+k_2+\eir}
\left(1+\frac{u}{t}\right)^{l}\nonumber \\
&\times&\biggl[\hyg\left(1+l,l+k_2+\eir,1+l+k_2+\eir,-\frac{\u}{\bar t}\right)
\nonumber \\
&-&\left(\frac{\tpo}{\s}\right)^{l+k_2+\eir}
\hyg\left(1+l,l+k_2+\eir,1+l+k_2+\eir,-\frac{\u \tpo}{{\bar t}\s}\right)
\biggr], 
\end{eqnarray}
In this case, the result might be IR finite. 
Then, after an $\eir$ expansion with the $\overline{MS}$ scheme,
it becomes
\begin{eqnarray}
J_4(s,t,p_1^2,0,0,0;n_x,n_y,n_z)
&\rightarrow&\frac{1}{(4 \pi)^2s~t}\left[
\Cz_{49}
+O(\eir)\right], 
\end{eqnarray}
where
\def\axyzmt{{\cal A}^{(-1)}_{n_x,n_y+n_z}(\t)}
\def\axyzzt{{\cal A}^{(0)}_{n_x,n_y+n_z}(\t)}
\def\axyzpt{{\cal A}^{(1)}_{n_x,n_y+n_z}(\t)}
\def\axyzms{{\cal A}^{(-1)}_{n_x,n_y+n_z}(\s)}
\def\axyzzs{{\cal A}^{(0)}_{n_x,n_y+n_z}(\s)}
\def\axyzps{{\cal A}^{(1)}_{n_x,n_y+n_z}(\s)}
\def\fmzzk{{\cal F}^{(-1)}_{0,0,k_2}}
\def\fzzzk{{\cal F}^{(0)}_{0,0,k_2}}
\def\fpzzk{{\cal F}^{(1)}_{0,0,k_2}}
\def\fmlzk{{\cal F}^{(-1)}_{l,0,k_2}}
\def\fzlzk{{\cal F}^{(0)}_{l,0,k_2}}
\def\fplzk{{\cal F}^{(1)}_{l,0,k_2}}
\def\fmxzy{{\cal F}^{(-1)}_{n_x,n_z,n_y}}
\def\fzxzy{{\cal F}^{(0)}_{n_x,n_z,n_y}}
\def\fpxzy{{\cal F}^{(1)}_{n_x,n_z,n_y}}
\begin{eqnarray}
\Cz_{49}&=&  
n_x\left(-\frac{t}{s}\right)^{n_x}\biggl[
\axyzzt\fpxzy\ess+\axyzpt\fzxzy\ess
\nonumber \\
&~&-{\cal H}_{n_x-1}\axyzzt\fzxzy\ess
\biggr]
\nonumber \\
&+&\sum_{k_1=0}^{n_z}\sum^{n_y+k_1}_{k_2=0}
~_{n_z}C_{k_1}~
~_{n_y+k_1}C_{k_2}~ 
(-1)^{k_1+k_2}
\left(\frac{s}{s-p_1^2}\right)^{n_y+k_1}
\nonumber \\
&~&\times\biggl[
\axyzzs
\left(\fzzzk\est-\fracp^{k_2}\left(\fzzzk\estt
+\lnaa\fmzzk\estt\right)
\right)
\nonumber \\
&~&+\axyzps
\left(\fmzzk\est-\fracp^{k_2}\fmzzk\estt\right)
\biggr]
\nonumber \\
&+&\sum_{l=1}^{n_x}l(-1)^l
~_{n_x}C_l~
\left(1+\frac{u}{t}\right)^l
\sum_{k_1=0}^{n_z}\sum_{k_2=0}^{n_y+k_1}
~_{n_z}C_{k_1}~
~_{n_y+k_1}C_{k_2}~ 
(-1)^{k_1+k_2}
\left(\frac{s}{s-p_1^2}\right)^{n_y+k_1}
\nonumber \\
&~&\times\biggl[
\axyzzs\left(
\fplzk\est-\fracp^{l+k_2}\left(\fplzk\estt
+\lnaa\fzlzk\estt\right)\right)
\nonumber \\
&~&
\left(\axyzps-\axyzzs{\cal H}_{l-1}\right)
\left(\fzlzk\est-\fracp^{l+k_2}\fzlzk\estt\right)
\biggr].
\label{res2}
\end{eqnarray}
\end{document}